\pdfoutput=1
\documentclass[a4paper,11pt]{article}

\usepackage{jcappub}

\usepackage[cp1250]{inputenc}
\usepackage{graphicx}
\usepackage{epstopdf}
\usepackage{amsmath,amsthm,latexsym,amssymb,amsfonts,epsfig}
\usepackage{fontenc,layout}
\usepackage{hyperref}
\usepackage[usenames,dvipsnames]{xcolor}

\newcommand{\be}{\begin{equation}}
\newcommand{\ee}{\end{equation}}
\newcommand{\bea}{\begin{eqnarray}}
\newcommand{\eea}{\end{eqnarray}}
\newcommand{\beq}{\begin{eqnarray}}
\newcommand{\eeq}{\end{eqnarray}}

\numberwithin{equation}{section}

\title{Intergalactic Magnetic Fields from First-Order Phase Transitions}

\author[1,2,3]{John~Ellis,}
\author[1]{Malcolm~Fairbairn,}
\author[1,4]{Marek~Lewicki,}
\author[1,3]{Ville~Vaskonen,}
\author[1]{Alastair~Wickens}
\affiliation[1]{Department of Physics, King's College London, Strand, WC2R 2LS London, UK}
\affiliation[2]{Theoretical Physics Department, CERN, Geneva, Switzerland}
\affiliation[3]{National Institute of Chemical Physics \& Biophysics, R\"avala 10, 10143 Tallinn, Estonia}
\affiliation[4]{Faculty of Physics, University of Warsaw ul.\ Pasteura 5, 02-093 Warsaw, Poland}
\emailAdd{john.ellis@cern.ch}
\emailAdd{malcolm.fairbairn@kcl.ac.uk}
\emailAdd{marek.lewicki@kcl.ac.uk}
\emailAdd{ville.vaskonen@kcl.ac.uk}
\emailAdd{alastair.wickens@kcl.ac.uk}

\abstract{We study the generation of intergalactic magnetic fields in two models for first-order phase transitions in the early Universe that have been studied previously in connection with the generation of gravitational waves (GWs): the Standard Model supplemented by an $|H|^6$ operator (SM+$H^6$) and a classically scale-invariant model with an extra gauged U(1) $B - L$ symmetry (SM$_{B-L}$). We consider contributions to magnetic fields generated by bubble collisions and by turbulence in the primordial plasma, and we consider the hypotheses that helicity is seeded in the gauge field or kinetically. We study the conditions under which the intergalactic magnetic fields generated may be larger than the lower bounds from blazar observations, and correlate them with the observability of GWs and possible collider signatures. In the SM+$H^6$ model bubble collisions alone cannot yield large enough magnetic fields, whereas turbulence may do so. In the SM$_{B-L}$ model bubble collisions and turbulence may both yield magnetic fields above the blazar bound unless the B$-$L gauge boson is very heavy. In both models there may be observable GW and collider signatures if sufficiently large magnetic fields are generated.
\\
\\
\\
\\
KCL-PH-TH/2019-60, CERN-TH-2019-104}

\begin{document}

\maketitle

\section{Introduction}

The existence, magnitude and origin of an intergalactic magnetic field (IGMF) have long been
topics for debate. Until relatively recently there were only upper limits
on its possible magnitude coming from Big-Bang Nucleosynthesis
(BBN) \cite{Grasso:2000wj,Kawasaki:2012va}, and measurements of the spectrum and anisotropies of the cosmic
microwave background (CMB) \cite{Seshadri:2009sy,Ade:2015cva,Jedamzik:1996wp,Jedamzik:1999bm,Barrow:1997mj,Durrer:1999bk,Yamazaki:2012pg,Trivedi:2010gi}. However, for some years now lower limits on the magnitude of
the IGMF have been reported \cite{Tavecchio:2010mk,Ando:2010rb,Neronov:1900zz,Essey:2010nd,Chen:2014rsa},
in particular in a recent analysis by the {\it Fermi}-LAT
collaboration~\cite{Biteau:2018tmv}, based on their observations of blazars in conjunction
with measurements of very-high-energy (VHE) $\gamma$ emissions by imaging
air {\v C}erenkov telescopes (IACTs). In combination, these provide evidence
for electromagnetic cascades interpreted as being due to the processes
of pair production via the IGMF: $\gamma + {\bf B} \to e^+ e^-$
followed by $\gamma$ emission during $e^\pm$ scattering off the magnetic field: $e^\pm + {\bf B} \to e^\pm + \gamma$.
The inferred magnitude of the IGMF depends on its unknown coherence
length $\lambda$ and on the unknown duration of blazar emissions.
Making the very conservative assumption that the blazars studied
have been active at a similar level for at least 10 years, 
the {\it Fermi}-LAT collaboration established the lower limit $|{\bf B}| >
3 \times 10^{-16}$~Gauss over a large range of $\lambda > 10^{-2}$~Mpc \cite{Biteau:2018tmv},
increasing at smaller $\lambda$, rising to $|{\bf B}| > 10^{-14}
(3 \times 10^{-13})$~Gauss for plausible active periods of $10^4 (10^7)$~years.

The origin of such a field on large scales does not yet have a satisfactory explanation. There are several astrophysical scenarios for the generation of magnetic fields on cluster and galaxy scales which involve battery effects creating seed fields \cite{1950ZNatA...5..237S,Kulsrud:1996km,Gnedin:2000ax} which are subsequently amplified to the observed strength in galaxies and clusters by dynamo action  \cite{1984JFM...144....1Z,Shukurov:2005pf,Kulsrud:2007an}. Such mechanisms have difficulty in explaining magnetic fields in the large voids which would be required to explain the constraints from gamma rays outlined above, so one is led to
envisage possible primordial sources originating from processes
involving particle physics. Natural sources include non-adiabatic
episodes in the early universe such as cosmological inflation \cite{Turner:1987bw,Ratra:1991bn,Martin:2007ue,Kobayashi:2014sga} or some
phase transition, e.g., the QCD or electroweak phase transition \cite{Vachaspati:1991nm,Sigl:1996dm,Tevzadze:2012kk}.
The QCD phase transition is well understood and thought to have been 
rather smooth, and hence unlikely to have generated large primordial
magnetic fields. The electroweak phase transition would also have been
quite smooth in the Standard Model (SM), but there is scope for
extensions of the SM that could have generated a first-order
phase transition that might have created significant primordial 
magnetic fields \cite{Espinosa:2011ax}.

A first-order electroweak transition would be interesting for 
several other reasons. For example, it could re-open the 
way to electroweak baryogenesis \cite{Kuzmin:1985mm}~\footnote{A 
primordial magnetic field could also explain the baryon asymmetry~\cite{Fujita:2016igl,Kamada:2016eeb,Kamada:2016cnb}.}
and could have sourced observable 
gravitational waves (GWs) \cite{Witten:1984rs,Turner:1987bw}. Moreover, the necessary extensions
of the SM might have observable signatures in particle collider
experiments, leading to the possibility of correlating laboratory
and GW measurements with measurements of the IGMF.

Here we study IGMF generation in two possible extensions of the SM that we have used previously to explore the possible magnitudes of GW signatures. One is the SM supplemented by an $|H|^6$ operator (SM+$H^6$)~\cite{Ellis:2018mja}, and the other is a classically scale-invariant
extension of the SM with an extra gauged U(1) $B - L$ symmetry 
(SM$_{B-L}$)~\cite{Ellis:2019oqb}. 

In exploring the possible generation of the IGMF in these models,
we consider two contributions arising from non-adiabatic processes
during a first-order phase transition: bubble collisions and
turbulence in the primordial plasma, were both found
to make significant contributions to the GW signals produced in these
two models~\cite{Ellis:2018mja,Ellis:2019oqb}. We also consider two different sources 
that lead to significant amplification of the magnetic field via an inverse cascade 
process: primordial helical field configurations \cite{Cornwall:1997ms,Giovannini:1997gp} and kinetic helicity \cite{1996PhRvE..53.1283S,1999PhRvL..83.3198J,1981PhRvL..47.1060M}. 

The outline of this paper is as follows. In Section~\ref{sec:magnetic-sources}
we review mechanism for generating a primordial magnetic field during a first-order phase transition,
in Section~\ref{evolution} we discuss the subsequent evolution of the primordial field
in a plasma background, and experimental constraints on the IGMF are reviewed in
Section~\ref{sec:experimentalconstraints}. We present illustrative results in the two
scenarios we sudy, SM+$H^6$ and SM$_{B-L}$ in Section~\ref{sec:results}, correlating
the calculated IGMF in these models with their possible GW and collider signals. Finally,
Section~\ref{sec:conx} summarizes our conclusions. We use natural units with $c = \hbar = k_b = \mu_0 = 1$.

\section{Primordial Magnetic Field Generation} \label{sec:magnetic-sources}

In this Section we discuss possible sources of magnetic fields generated during a 
first-order phase transition, using an approach similar to that 
used previously~\cite{Ellis:2018mja,Ellis:2019oqb} for 
gravitational wave production. We expect that some fraction of the energy released 
during the phase transition would generically be available for the production of 
magnetic fields, and can divide the sources into two main sub-classes:
\begin{itemize}
    \item energy stored in the bubble walls, which source $\bf{B}$-fields upon collision,
    \item turbulent kinetic energy in the charged plasma that is
    available for later $\bf{B}$-field generation via magnetohydrodynamic (MHD) mechanisms.
\end{itemize}

 At any given time $t \geq t_*$ after the phase transition, we can describe 
 both the magnetic field and the turbulent plasma using their respective energy density spectra,
 $\rho_i(\lambda, t) = {\rm d}\rho_i(t)/{\rm d}\log\lambda$, $i=B,K$,
 and the corresponding magnetic field spectrum is then given by 
\be \label{Brho}
B(\lambda,t) = \sqrt{2\rho_B(\lambda,t)}\,.
\ee
The mean energy densities of the two components at this time are
$\rho_{i}(t) = \int \rho_i(\lambda,t) {\rm d}\log\lambda$. We define $v^2_i(t) = 2\rho_i(t)/\rho_0$,
so that $v_B$ and $v_K$ are, respectively, the Alfv\`en velocity associated with the magnetic field 
and the root-mean-square plasma velocity, while $\rho_0$ describes the initial plasma energy density.

Magnetic fields sourced from bubble collisions were recently revisited in~\cite{Zhang:2019vsb},
where it was found that about $10\%$ of the energy of the transition was expended on production 
of magnetic fields, mostly as the field oscillates around the minimum of the potential after the 
transition has been completed. In this simulation all the energy from vacuum conversion was 
used to accelerate the bubble walls. In order to use these results in the more realistic 
setting of a transition taking place in a plasma-filled background,
we need to include the appropriate efficiency factor 
\be
\kappa_{\rm col} =\frac{\rho_{\rm wall}}{\rho_{\rm V}} \, ,
\ee
which describes the fraction of energy used to accelerate the bubbles,
just as in case of GW generation (see~\cite{Ellis:2019oqb} for details). 

For the plasma-related sources the efficiency factor is
\be
\kappa_{\rm sw} =\frac{\alpha_{\rm eff}}{\alpha} \frac{\alpha_{\rm eff}}{0.73+0.083\sqrt{\alpha_{\rm eff}}+\alpha_{\rm eff}} \, , \quad {\rm with} \quad \alpha_{\rm eff} = \alpha(1-\kappa_{\rm col}) \,,
\ee 
which describes fraction of the energy converted into bulk fluid motion~\cite{Espinosa:2010hh,Caprini:2015zlo}, including also through $\alpha_{\rm eff}$ 
the fraction of energy used for bubble acceleration~\cite{Ellis:2019oqb}. Moreover, 
we assume that the efficiency for converting bulk fluid motion of the plasma into magnetic fields 
via MHD turbulence is $10\%$~\cite{Kahniashvili:2009qi,Durrer:2013pga,Brandenburg:2017neh}. 
This value is an order-of-magnitude estimate at best, but
it will be very simple to re-scale our estimates below with more accurate results
once more detailed estimates are available. This assumption
leads to a simple final expression for the energy density of the magnetic field 
after the phase transition:
\be \label{eq:rhoBstar}
\rho_{B,*} = \frac{0.1 \kappa\,\alpha}{1+\alpha} \rho_* \,,
\ee
where $\kappa$ is the efficiency factor for the source in question 
(either $\kappa_{\rm col}$ or $\kappa_{\rm sw}$ ) and $ \rho_*=3 M^2_p H_*^2$ 
is the total energy density at the time of percolation. 

Taking the Fourier transform of the mean magnetic and kinetic energy densities, 
we obtain a power spectrum for each of these components $P_B(\lambda)$ and $P_K(\lambda)$. 
These power spectra can then be expressed in terms of their associated energy density spectra as
\begin{equation} \label{power-spectra}
  P_{i}(\lambda,t) = \frac{\lambda^3}{4\pi}   \rho_i(\lambda, t),
\end{equation}
where $i=B,K$ describe quantities related to the magnetic and turbulent plasma fields respectively. 
We can represent the initial configuration of the two fields in terms of their power spectra, 
both of which peak at the average bubble size at collision, $\lambda_* = R_*$:
\begin{equation} \label{initial-power}
    P_{*,i}(\lambda, t_*) \approx P_{*,i}\lambda^{-n_i(\lambda)}
\end{equation}
where we have approximated each initial power spectrum with a separate power law. 

At scales $\lambda>\lambda_*$, causality and the divergence-free condition on
the initial magnetic field require that $n_B \geq 2$ \cite{Durrer:2003ja}, 
whilst in the case of the plasma the requirement is less stringent,
with $n_K \geq 0$, where $n_K=0$ and $2$ correspond to compressible and incompressible plasmas,
respectively, both of which are feasible scenarios in an MHD system.

We do not consider the slope of the initial spectrum at length scales smaller than $\lambda_*$, 
as we generically expect turbulence, driven by the turbulent eddies at the coherence scale, 
to fully develop relatively quickly on these smaller length scales,
due to the smaller eddy turnover times involved. Once the turbulent eddies have developed 
on all scales $\lambda < \lambda_*$, magnetic and kinetic energy is subsequently transferred 
to increasingly smaller scales via a direct energy cascade~\footnote{This direct energy cascade 
arises due to the non-linear terms in the fluid equations that facilitate interactions between 
the different wavelength modes.}, until the energy is eventually lost as heat at the 
dissipation scale of the plasma, $\lambda_d$. When a turbulent eddy develops on a particular 
distance scale, we expect the power spectrum of the initial magnetic or plasma field 
to be `processed' into an independent spectrum following a global turbulent decay law. 
In our work we assume that the form of any processed turbulent spectrum follows
a Kolmogorov decay law \cite{Kolmogorov:1941}  with $n_i=-11/3$,
as argued on dimensional grounds in \cite{Landau1987Fluid}. 

Assuming the power $n_B = 2$ at large scales, the initial magnetic field energy density 
spectrum is then given by
\be \label{rhoBspectrum}
\rho_B(\lambda,t_*) = \frac{17}{10} \,\rho_{B,*} \,
\begin{cases}
\left(\frac{\lambda}{\lambda_*}\right)^{-5}  & \mbox{ for }~~ \lambda\geq\lambda_* \,, \\    \left(\frac{\lambda}{\lambda_*}\right)^{2/3} & \mbox{ for }~~ \lambda < \lambda_* \,.
\end{cases}
\ee
In the following we discuss how this spectrum evolves. The evolution changes only the
amplitude and the coherence scale of the spectrum, but does not change its shape.

\section{Evolution of the Magnetic Field in a Plasma Background}
\label{evolution}

The initial power in the magnetic field subsequently spreads over a variety of 
different length scales, due to the non-zero coupling between the magnetic and plasma components. 
To study the evolution of the magnetic field in the turbulent plasma we need to track 
how key quantities, such as the comoving magnetic coherence length, $\lambda_B$, 
and the comoving magnetic field strength, $B$, change with time from their initial values 
set immediately after the phase transition completes, at $t=t_*$. 
Properties related to both the turbulent plasma and inherent to the magnetic field itself, 
e.g., non-zero magnetic helicity, can significantly impact this evolution,
as we discuss below, broadly following the approach in~\cite{Durrer:2013pga}. 

It is important to emphasize that the scaling laws of the magnetic field quoted in this 
section apply only in the presence of a plasma background. As such, we
assume they are relevant only until the time of recombination, $t_{\textrm{rec}}$, 
and that subsequently the magnetic field strength simply dilutes as normal due to redshift, 
$B \sim a^{-2}$. 

Immediately after the phase transition, turbulent eddies have only had time to develop
fully~\footnote{By `develop fully' we mean that the plasma eddy needs to have completed 
one rotation on the scale of interest.} on scales $\lambda<\lambda_*$. 
At subsequent times after the transition, turbulent eddies begin to develop on 
increasingly larger scales $\lambda>\lambda_*$, and the power spectrum on these 
scales is processed from its initial slope to one characterised by a Kolmogorov decay law. 
Thus the maximum comoving distance on which the field is correlated, known as the comoving 
coherence scale, $\lambda_i$, increases with time. We can approximate the time evolution of 
$\lambda_i$ by of a power law
\begin{equation}
    \lambda_i \sim t^{\zeta_i}\,,  \quad \quad i=B,K\,,
\end{equation}
where $\zeta_i >0$. Assuming Kolmogorov turbulent decay, the direct energy cascade 
transfers both magnetic energy and turbulent plasma energy to increasingly smaller scales 
before it is eventually dissipated into heat. Thus the comoving energy density of both the 
magnetic field, $\rho_B$, and the turbulent plasma, $\rho_K$, can be assumed to decay with time.

These energy densities define characteristic velocities for the respective magnetic and turbulent 
plasma field components, describing the rate at which changes in the 
relevant field can propagate. We can subsequently use these velocities
to describe the maximum length scale on which each of the fields are correlated 
at any given time $t\gg t_*$ after the transition:
\begin{equation}
    \lambda_i \sim v_i t\,.
\end{equation}
Once MHD turbulence has fully developed at a given scale, we expect equipartition 
between the plasma and magnetic energy densities, $\rho_B \sim \rho_K \sim \rho_{\rm eq}$. 
Using Eq.~\eqref{power-spectra} in conjunction with Eq.~\eqref{initial-power}, we can write
\begin{equation} \label{v_eq}
    v_{\rm eq} \sim \sqrt{2 \rho_{\rm eq}} \sim \lambda_{\rm eq}^{-\frac{3+n}{2}} \, ,
\end{equation}
where $\lambda_{\rm eq} \sim \lambda_B \sim \lambda_K$ and $v_{\rm eq}\sim v_B \sim v_K$ 
from the equipartition condition, and $n=\text{min}(n_B,n_K)$ corresponds to the 
spectral decay law of the field that initially dominates, whether it be the 
magnetic field or the turbulent plasma. The size of the largest `processed' eddy 
can then be expressed using Eq.~\eqref{v_eq} as
\begin{equation} \label{lamda_eq}
    \lambda_{\rm eq} \sim v_{\rm eq}t \sim t^{\frac{2}{5+n}} \,.
\end{equation}
Using Eqs.~\eqref{Brho} and \eqref{lamda_eq} we obtain the following scaling
laws based on dimensional arguments:
\begin{equation}  \label{B_vs_t}
B(\lambda_{\rm eq}) \sim \sqrt{2\rho_{\rm eq}} \sim t^{\frac{3+n}{5+n}} \,,
\end{equation}
and
\begin{equation} \label{B_vs_lambda}
B(\lambda_{\rm eq}) \sim \lambda_{\rm eq}^{-\frac{3+n}{2}} \, .
\end{equation}

\subsection{Helicity and Inverse Cascades}

As shown in~\cite{Biskamp:1999,biskamp_2003}, the behaviour of the field evolution can be 
changed dramatically in the presence of non-zero average magnetic helicity, 
$\langle \textbf{A} \cdotp \textbf{B} \rangle$, where $\textbf{B}=\bf{\nabla} \times \textbf{A}$, 
a quantity describing the twisting of magnetic field lines. We expect some fraction of 
magnetic helicity to be left over after the phase transition from baryon-number-violating 
processes such as decaying non-perturbative field configurations, 
e.g., electroweak sphalerons~\cite{Vachaspati:2001nb}. Furthermore, we expect some amount of 
helicity to be generated as the swirling plasma results in twisted field lines, 
due to the coupling between magnetic field and plasma components in an MHD system~\cite{biskamp_2003}.

In a highly conductive plasma we expect magnetic helicity to be conserved:
$\langle \textbf{A} \cdotp \textbf{B} \rangle \propto \rho_B\lambda_B = \text{const}$,
and thus $B \sim \sqrt{\rho_B} \sim 1/\sqrt{\lambda_B}$, which we can use with 
Eq.~\eqref{B_vs_lambda} to deduce that this corresponds to the case $n=-2$. 
Thus from Eq.~\eqref{lamda_eq} and \eqref{B_vs_t} the time evolution of the 
comoving coherence scale $\lambda_B$ and the comoving magnetic field strength $B$ 
in the presence of magnetic helicity can be argued to be~\cite{Biskamp:1999}
\be \label{helicalscaling}
    \lambda_{\rm eq} \sim t^{\frac{2}{3}} \,,\qquad
    B(\lambda_{\rm eq}) \sim t^{-\frac{1}{3}} \,.
\ee
Assuming some initial fractional magnetic helicity immediately after the transition, 
we expect the field to reach a maximally helical state some time later~\cite{Tevzadze:2012kk},
due to the direct energy cascade, which results in the conversion of the large-scale 
non-helical field component to heat~\cite{Jedamzik:1996wp}.

More recently, numerical simulations in~\cite{Brandenburg:2017rnt} have shown that 
inverse cascade behaviour also takes place when a non-helical magnetic field is in the 
presence of a plasma with initial kinetic helicity. The scalings of the comoving 
coherence scale and comoving magnetic field amplitude with time in this scenario 
are then found to be
\be \label{kineticscaling}
    \lambda_{\rm eq} \sim t^{\frac{1}{2}} \,, \qquad    
    B(\lambda_{\rm eq}) \sim t^{-\frac{1}{2}} \,.
\ee

\subsection{Magnetic Field Spectrum Today}

Using the above results for the evolution of the comoving magnetic field spectrum, 
we can calculate the spectrum today. The scaling laws~\eqref{helicalscaling} 
and \eqref{kineticscaling} hold until recombination, which happens only shortly 
after matter-radiation equality, so it is a good approximation that the scale factor behaves as $a\sim t$.
So, redshifting the initial spectrum~\eqref{rhoBspectrum} while assuming a decay law $\sim a^{-q_B}$ of the comoving magnetic field until recombination gives the following magnetic field spectrum today:
\be
B_0(\lambda) \equiv B(\lambda,t_0) = \left(\frac{a_*}{a_{\rm rec}}\right)^{q_B} \left(\frac{a_*}{a_0}\right)^2 \sqrt{\frac{17}{10}\,\rho_{B,*}} \,
\begin{cases}
\left(\frac{\lambda}{\lambda_0}\right)^{-5/2}  & \mbox{ for }~~ \lambda\geq\lambda_0 \,, \\    \left(\frac{\lambda}{\lambda_0}\right)^{1/3} & \mbox{ for }~~ \lambda < \lambda_0 \,,
\end{cases}
\ee 
where the initial magnetic field energy density $\rho_{B,*}$ is given by Eq.~\eqref{eq:rhoBstar}. Similarly, assuming that the comoving field coherence scale evolves as 
$\sim a^{q_\lambda}$ until recombination, the field coherence scale today is
\be
\lambda_0 \equiv \lambda_B(t_0) = \left(\frac{a_{\rm rec}}{a_*}\right)^{q_\lambda} \left(\frac{a_0}{a_*}\right) \lambda_* \,,
\label{peakscale}
\ee
where the initial coherence scale is given by the bubble size at percolation, $\lambda_* = R_*$.  Assuming fast reheating to a temperature $T_{\rm reh}$ after the phase transition,
the factor $a_*/a$, where $a$ is either $a_{\rm rec}$ or $a_0$, can be expressed as
\be
\frac{a_*}{a} = \left(\frac{h_{\rm eff}(T)}{h_{\rm eff}(T_{\rm reh})}\right)^\frac13 \frac{T}{T_{\rm reh}} \,,
\ee
where $h_{\rm eff}(T)$ is the effective number of entropy degrees of freedom at temperature $T$.

\section{Experimental Constraints on the IGMF}
\label{sec:experimentalconstraints}

As mentioned in the Introduction, very-high-energy gamma rays 
emitted by TeV blazars will Thompson scatter 
with the photons in the Extragalatic Background Light (EBL) to produce 
highly-relativistic $e^{+}e^{-}$ pairs. These pairs are subsequently 
expected to lose energy via inverse Compton scattering with CMB photons, 
finally producing a significant emission of secondary cascade photons 
in the GeV band. 

A lower bound on the IGMF can be inferred from the non-observation 
of such GeV cascade photons. In the presence of an IGMF, $e^{+}e^{-}$ pairs produced during the cascade process would be deflected from the trajectory expected in the absence of the magnetic field. Thus the total emission in the 
GeV band would be reduced for increasing IGMF strength, as the initial 
incoming photon flux is distributed over a greater solid angle. 
Additionally, a stronger IGMF would also result in an increasing 
time delay between the cascade GeV and direct TeV emissions. Both these 
effects provide mechanisms for suppressing the flux of GeV cascade photons, 
which can then be re-expressed in terms of a lower bound on the strength of the IGMF.

Refs.~\cite{Neronov:1900zz,Tavecchio:2010mk,Dolag:2010ni, Ando:2010rb,Biteau:2018tmv} analysed {\it Fermi} data from blazars possessing hard TeV spectra with negligible GeV cascade emission, under the assumption that this suppression was due to the large angular size of the cascade emission. They each derived a constraint, later verified by~\cite{Taylor:2011bn} using simultaneous GeV/TeV energy band observation data, on the minimum value of the IGMF. The constraints reported varied in the range $B_0 \gtrsim 10^{-16} - 10^{-15}$\,G for $\lambda_0 \gtrsim 1 $\,Mpc depending on the adopted EBL and source model. For our purposes we adopt the {\it Fermi}-LAT constraint in~\cite{Biteau:2018tmv}, $B_0> 3\times 10^{-16}$\,G, corresponding to the blue solid line of the plots in Section~\ref{sec:results}. Their analysis shows that the high-latitude sources detected by the {\it Fermi}-LAT do not have significant spatial extension.

Independently, Refs.~\cite{Dermer:2010mm,Taylor:2011bn} considered 
explaining the suppressed GeV emission by an IGMF-induced time delay 
in the GeV cascade signal compared to the TeV direct signal. 
Using simultaneous multi-band observation data from {\it Fermi} and 
ground-based Cherenkov telescopes, and assuming that the blazars 
had not been firing for a long period, so that the cascade signal 
had not yet reached the detectors, these authors were able to deduce 
lower bounds on the IGMF lying in the range $B_0 \gtrsim 10^{-18} - 10^{-17}$\,G for $\lambda_0 \gtrsim 1$\,Mpc. The difference between the values obtained by the two investigations can be explained by the method used to model the cascade signal, and in our work we take the constraint from~\cite{Taylor:2011bn}, $B_0 > 10^{-17}$\,G, shown as the dashed blue line in the plots in Section~\ref{sec:results}.

The above constraints on the IGMF are independent of the magnetic coherence 
scale for $\lambda_0 \gtrsim 1$\,Mpc, because the length scales on which 
$e^{+}e^{-}$ pairs undergo inverse Compton cooling are much smaller than 
those over which the underlying field is correlated. In this scenario 
the pairs can be approximated as moving through a homogeneous magnetic field.
However, for scales $\lambda_0 \lesssim 1$\,Mpc this simplifying 
approximation is no longer valid and the $e^{+}e^{-}$ pairs may 
experience changes in their deflection trajectory as they journey through
uncorrelated patches of the IGMF. Thus an uncertainty in the final deflection 
angle and time delay of the signal is introduced at these scales. 
To account for this, the strength of the magnetic field should increase 
like $B_0 \sim \lambda_0^{-1/2}$ at scales $\lambda_0 \lesssim 1$\,Mpc, 
in order to ensure that the trajectories of the $e^{+}e^{-}$ pairs are 
deflected to the degree required to explain the absence of the GeV 
emission in the experimental data.

For our plots in Section~\ref{sec:results} we take the IGMF constraint associated with the extended angular cascade emission from Ref.~\cite{Biteau:2018tmv}, where the bound is taken to be flat for $\lambda_0 \gtrsim 10$\,kpc and increases like $B_0 \sim \lambda_0^{-1/2}$ below such scales. On the other hand, for the IGMF constraint associated with the time delay of the cascade signal we take the result from Ref.~\cite{Dolag:2010ni}, where the bound is taken to be flat for $\lambda_0 \gtrsim 100$\,kpc.

\section{Results in Illustrative Examples}
\label{sec:results}

In order to calculate the IGMF generated by a particular phase transition, 
we need to be more specific about the scenario that we consider.  
Here we follow closely the two scenarios explored in~\cite{Ellis:2019oqb}, 
the first being a Higgs sector with an additional non-renormalizable
$|H|^6$ operator (SM+$H^6$), and the second being a classically 
scale-invariant extension of the SM with an extra gauged 
U(1) $B - L$ symmetry (SM$_{B-L}$).

The generation of GWs in these models has been analyzed previously in~\cite{Ellis:2018mja,Ellis:2019oqb} and we follow the procedure described there in the current analysis. 
The GW signal produced by turbulence could be enhanced due to the inverse cascade occurring during the evolution of MHD turbulence with a helical magnetic field~\cite{Kahniashvili:2008pe}. The impact of this modification, however, depends on the modelling of the evolution of the turbulence and will not change the experimental reach into the models at hand, as turbulence is usually not the dominant source of GWs. Still, observation of the tail of the signal produced by turbulence could probe the helicity of the source through polarisation of the GW signal~\cite{Kisslinger:2015hua}.

\subsection{Standard Model with $\left|H\right|^6$ term}
\label{sec:H6}

\begin{figure}
    \centering
    \includegraphics[width=0.99\textwidth]{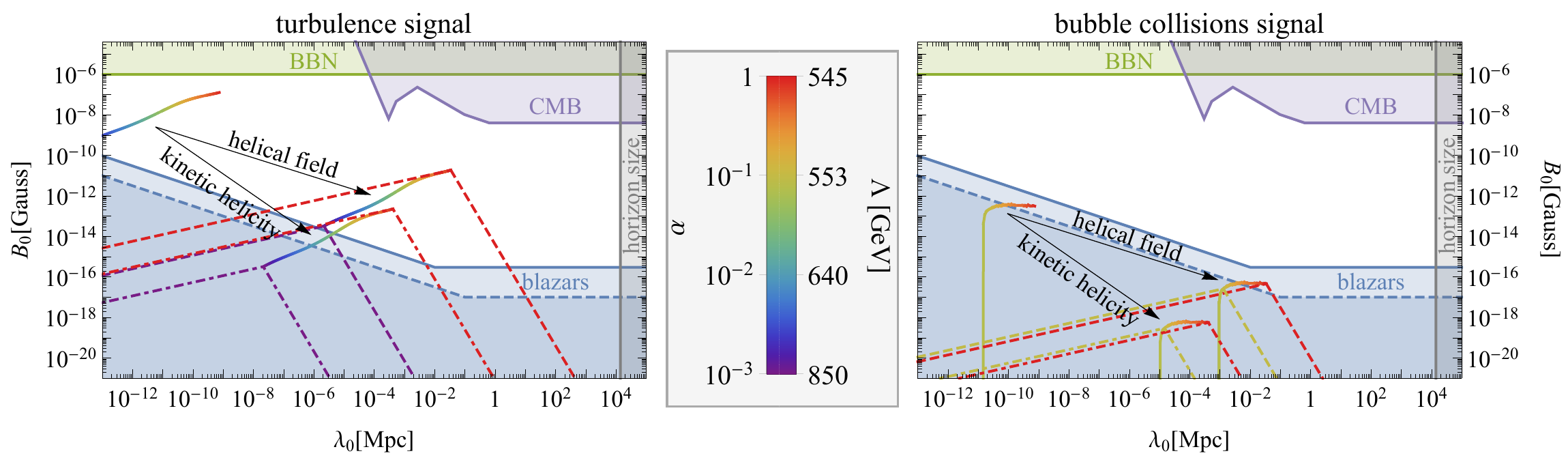}
    \caption{The magnetic field strength $B_0$ as a function of its coherence length 
    $\lambda_0$ produced by a phase transition in the SM$+H^6$ model. The colour coding indicates 
    the strength of the phase transition corresponding to a given cut-off scale $\Lambda$,
    while the arrows indicate the evolution of the field due to the inverse cascade in the plasma background. 
    The dashed lines show examples of final magnetic field spectra for a few selected transitions.
    The left panel shows the possible contribution of turbulence in the plasma, while the right panel 
    shows that due to bubble collisions.}
    \label{fig:SMrainbowPlot}
\end{figure}

This model serves as a minimal and natural extension of the SM in which to explore the consequences of a possible first-order electroweak phase transition for the intergalactic magnetic field. The dynamics of the phase transition in this model is quite generic, and it has features typical of many simple modifications of the SM featuring polynomial potentials as might be generated, for instance,
by a new neutral scalar. Specifically, the field cannot remain in the metastable vacuum too long 
without spoiling percolation. As a result, the strength of the phase transition is bounded by 
$\alpha \lesssim 1$ for all feasible transitions in this model. Large couplings with the 
plasma and the impossibility of cooling in order to decrease the plasma friction mean that the 
bubbles reach their terminal velocity very quickly, and most of the energy of the transition
is dumped into plasma shells surrounding the bubbles. In principle, these models could be mapped onto the recent simulations of first-order SM phase transitions~\cite{Zhang:2019vsb}. However, in those simulations, interaction of the wall with the plasma background was neglected and the source of the magnetic field was oscillations of the scalar field.  Since we find that most of the energy is transferred into the plasma, we would only expect a very small residual magnetic field to be produced - the fraction of energy transferred directly into the bubble walls is very small $\kappa_{\rm col}\approx 10^{-10}$ ~\cite{Ellis:2019oqb}. Instead, just as in the GW
case, the bulk of our results come from plasma-related sources, specifically turbulence 
developing in the plasma after the transition. The sound-wave period in these models 
lasts less than a Hubble time~\cite{Ellis:2018mja}, and the nonlinear dynamics following it 
may lead to a significant amount of bulk fluid motion which could be converted into turbulence~\cite{Caprini:2015zlo,Ellis:2019oqb}.

\begin{figure}
    \centering
    \includegraphics[width=7.3cm]{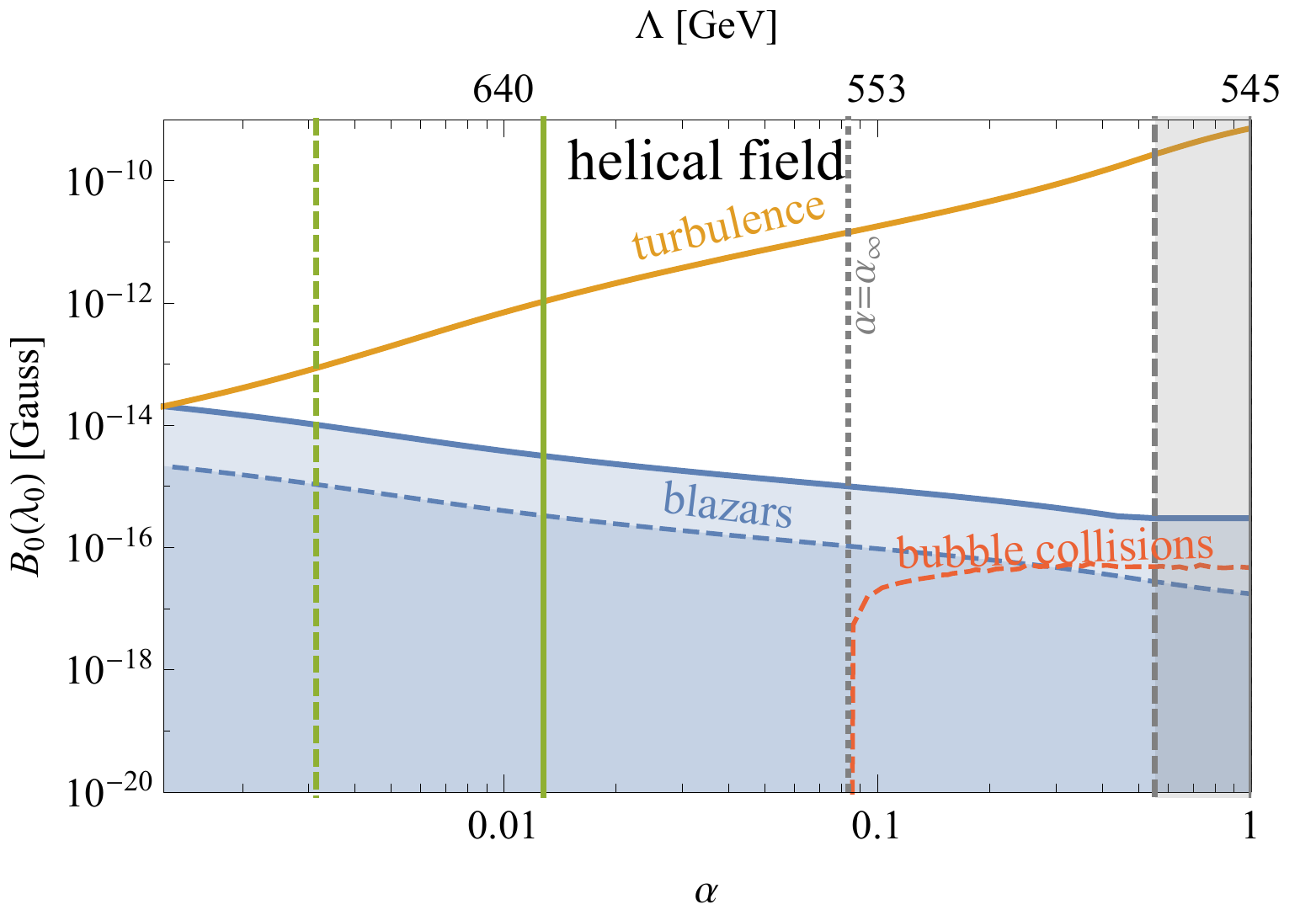}
     \includegraphics[width=7.3cm]{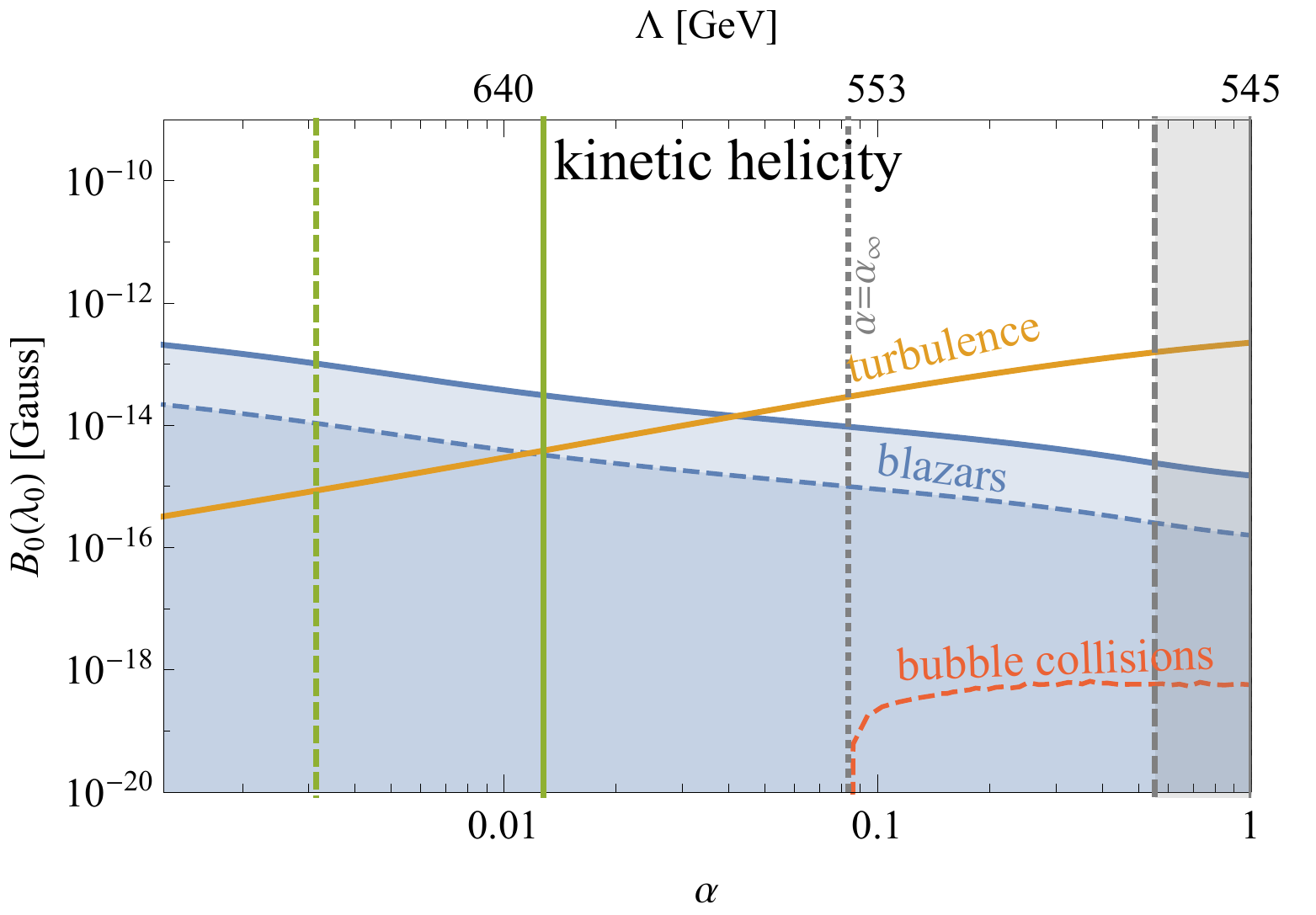}
    \caption{The red and yellow show the magnetic field strength $B_0$ at the coherence scale $\lambda_0$ as a function of the strength of the phase transition $\alpha$ sourced by the turbulence and bubble collisions, respectively, from the phase transition in the SM+$H^6$ model. We also show the corresponding $|H|^6$ operator scale $\Lambda$, where the solid (dashed) green lines indicate the 3- (2-)$\sigma$ sensitivities of HL-LHC. The left panel corresponds to evolution assuming a helical magnetic field while the right panel shows results assuming only kinetic helicity.  The blue shaded region shows the lower limit from blazar observations on the magnetic field at the coherence scale $\lambda_0$. The gray shaded region corresponds to where the percolation calculation becomes unreliable~\cite{Ellis:2018mja}.  The line $\alpha = \alpha_\infty$ corresponds to the value of $\alpha$ above which one overcomes the leading-order plasma friction.}
    \label{fig:SMexpPlot1}
\end{figure}

We show the results for both bubble collisions and plasma-related sources for all of the
SM$+H^6$ parameter space with a reasonably strong transition in Figs.~\ref{fig:SMrainbowPlot} and \ref{fig:SMexpPlot1}. The blue shaded regions show the lower limits on the value of the 
IGMF based on blazar observations discussed in Section~\ref{sec:experimentalconstraints}. 
We also show in Fig.~\ref{fig:SMrainbowPlot} the experimental constraints from BBN~\cite{Grasso:2000wj,Kawasaki:2012va} and the CMB~\cite{Ade:2015cva,Seshadri:2009sy, Jedamzik:1996wp,Jedamzik:1999bm,Barrow:1997mj,Durrer:1999bk,Yamazaki:2012pg,Trivedi:2010gi}~\footnote{One can also put constraints on much smaller field coherence scales through the GW
background produced by the magnetic field~\cite{Saga:2018ont}}. 
We see that the calculated field strength exhibits generically a peak at a
coherence scale $\lambda_0$ given by (\ref{peakscale}), which originates from the
break in the initial magnetic field energy density shown in (\ref{rhoBspectrum}). The
value of $\lambda_0$ at the peak and the corresponding value of $B_0$ depend on
the strength $\alpha$ of the transition, which depends in turn on the scale $\Lambda$
of the $|H|^6$ operator, as indicated by the colour coding. The left panel of
Figs.~\ref{fig:SMrainbowPlot} shows the possible contribution to the magnetic
field spectrum from turbulence, and the right panel shows that from bubble collisions.
In each panel we also exhibit the evolution of the field due to the inverse cascade
in the plasma background under the hypotheses that the primordial field is due to
helical field configurations or kinetic helicity.

We see that the magnetic field strength due to turbulence may well
exceed the blazar lower limit, peaking at a coherence scale $\sim 10^{-5}$ 
to $\sim 10^{-2}$ Mpc. On the other hand, it is not possible to explain
the blazar data with a magnetic field produced by bubble collisions in the plasma after the phase transition in this model, even when the transition is rather 
strong and the field becomes fully helical quickly after it is produced.

\begin{figure}
    \centering
    \includegraphics[width=10cm]{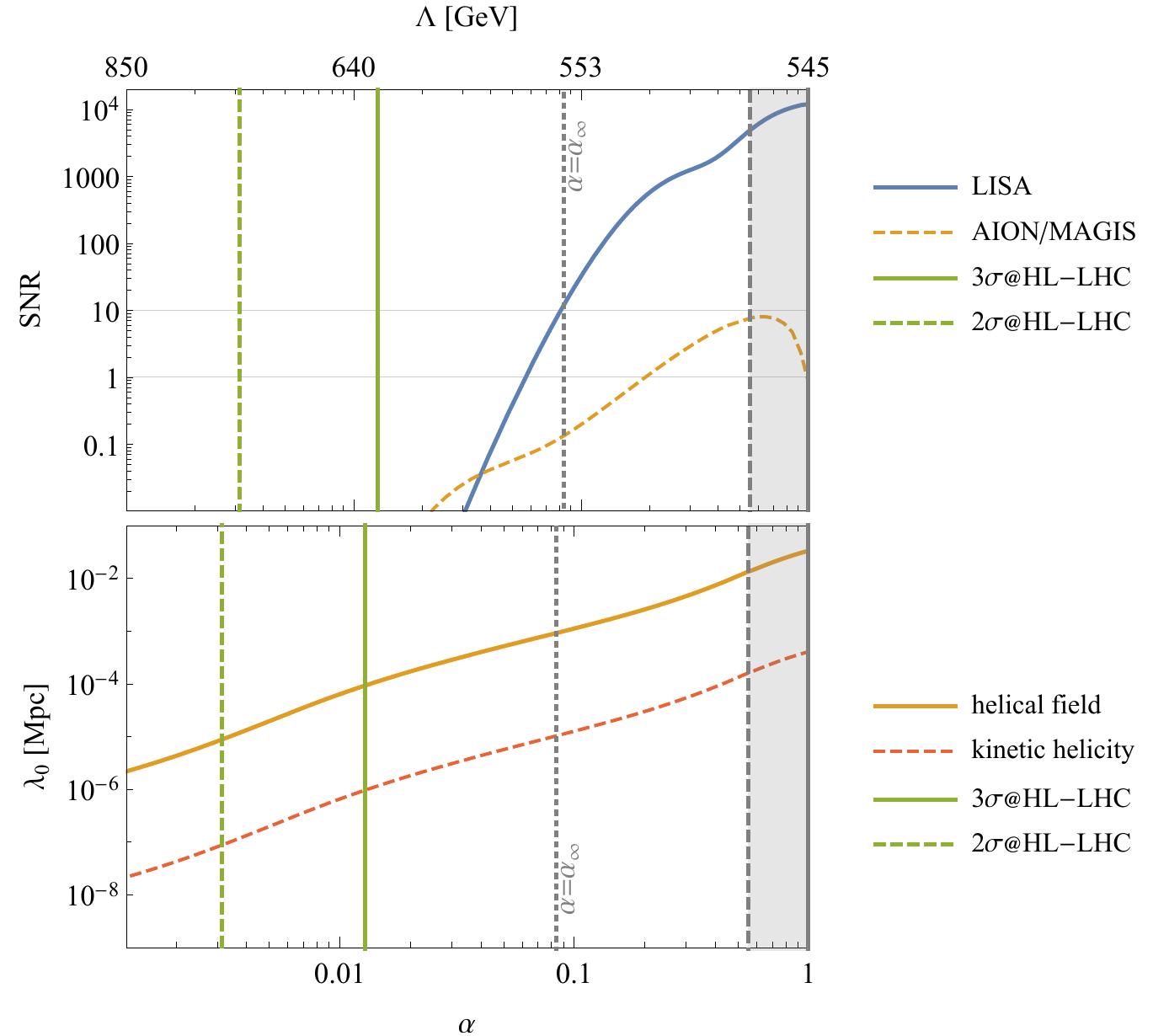}
    \caption{The top panel shows SNR of the GW signal from the phase transition in the SM+$H^6$ model for different experiments, while the lower panel indicates the magnetic field coherence scale. The solid and dashed lines in the lower panel correspond to evolution assuming a helical magnetic field or kinetic helicity, as shown in Fig.~\ref{fig:SMrainbowPlot}.}
    \label{fig:SMexpPlot2}
\end{figure}

The parameter region where the phase transition is strong offers other possible experimental probes. 
We consider first the GWs produced by the same phenomena responsible for magnetic field production~\cite{Huang:2015izx,Artymowski:2016tme,Chala:2018ari}. To illustrate this,
we show in Fig.~\ref{fig:SMexpPlot1} the strength of the magnetic field and in Fig.~\ref{fig:SMexpPlot2} the signal-to-noise ratio (SNR) in the planned GW experiments 
most relevant for this model LISA~\cite{Bartolo:2016ami}, MAGIS~\cite{Graham:2016plp,Graham:2017pmn} 
and AION~\cite{AION:2018} (see~\cite{Ellis:2019oqb} for more details on the GW spectra). As we see the strength of both  magnetic field and GW signal produced grows with the strength of the transition. Both the signals are also produced predominantly by plasma related sources with bubble collisions predicting a contribution too weak to explain the observed IGMF.
The conclusion here is that, as expected, the GW signals and magnetic field production are correlated and, in case of purely kinetic helicity, future GW experiments will be able to probe all of the parameter space in which the resulting magnetic field is strong enough to satisfy the blazar bounds. However, this is not necessarily the case if the field becomes fully helical at some point. 

Another promising avenue for testing such scenarios is in collider experiments, 
although here the details depend a lot more on the underlying particle physics model~\cite{Bodeker:2004ws,Delaunay:2007wb,Curtin:2014jma}. In our particular case of a
single non-renormalisable $H^6$ operator the only such probe is through the modification of the 
triple-Higgs coupling~\cite{Ellis:2018mja}, and we indicate the reach of HL-LHC in Fig~\ref{fig:SMexpPlot1} and Fig~\ref{fig:SMexpPlot2}, expressed as the scale $\Lambda$ associated with the $|H|^6$ operator. 
We see that HL-LHC will also be able to probe all of the parameter space relevant from the point of view of magnetic field production. 
However, this is rather a model-dependent statement that would not necessarily be true in the
next simplest extension, which is the SM with a singlet scalar~\cite{Ashoorioon:2009nf,Beniwal:2017eik}.

\subsection{Classically scale-invariant ${\rm U}(1)_{B-L}$ model}
\label{sec:csi}

Next we consider electroweak symmetry breaking in the classically scale-invariant U$(1)_{B-L}$ extension of the SM, in which case the transition can be very strong, $\alpha \gg 1$. The details of the model can be found, e.g., in Refs.~\cite{Iso:2009ss,Marzo:2018nov}. The breaking of the U$(1)_{B-L}$ gauge symmetry is triggered by a new scalar field $\varphi$. The thermal corrections to its scalar potential induce a barrier between the $B-L$-symmetric and $B-L$-breaking minima, so that the transition is first-order. These corrections are dominated by the $B-L$ gauge boson $Z'$. Therefore, the $B-L$ gauge coupling $g_{B-L}$ and the $Z'$ mass $m_{Z'}$ determine the dynamics of the transition. As the $\varphi$ field acquires a non-zero vacuum expectation value, it triggers electroweak symmetry breaking via the portal coupling $-\lambda_p h^2 \varphi^2/4$.

\begin{figure}
    \centering
    \includegraphics[width=0.99\textwidth]{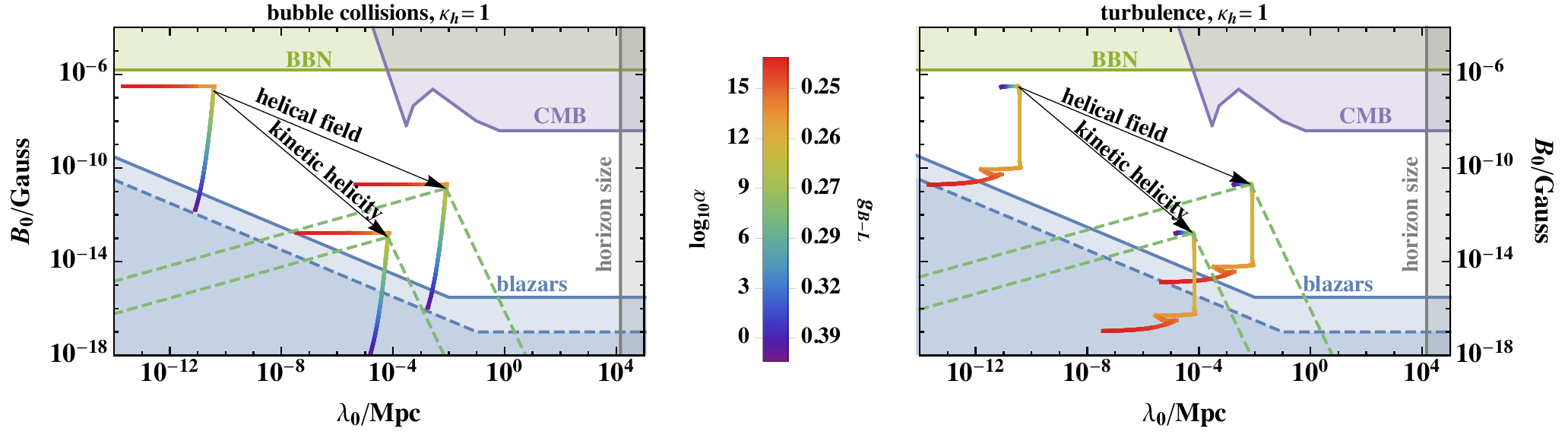} \\ \vspace{2mm}
    \includegraphics[width=0.99\textwidth]{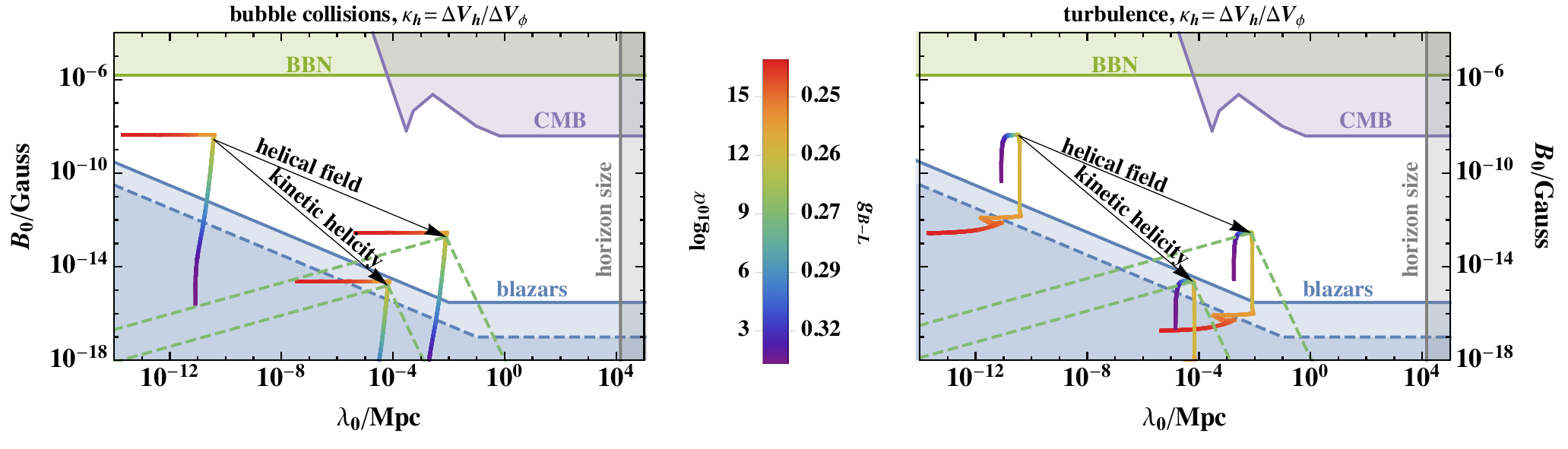}
    \caption{Results as in Fig.~\ref{fig:SMrainbowPlot} computed for $SM_{B-L}$ model with 
    $m_{Z'}=4\,{\rm TeV}$. The upper panels show the optimistic result assuming that all the phase transition energy is transferred into the visible sector, and the lower panels are for the 
    realistic case in which only the Higgs field bubbles produce magnetic fields, with no contribution from heavy fields beyond the SM.}
    \label{fig:Bplotcsi}
\end{figure}

\begin{figure}
    \centering
    \includegraphics[height=9.cm]{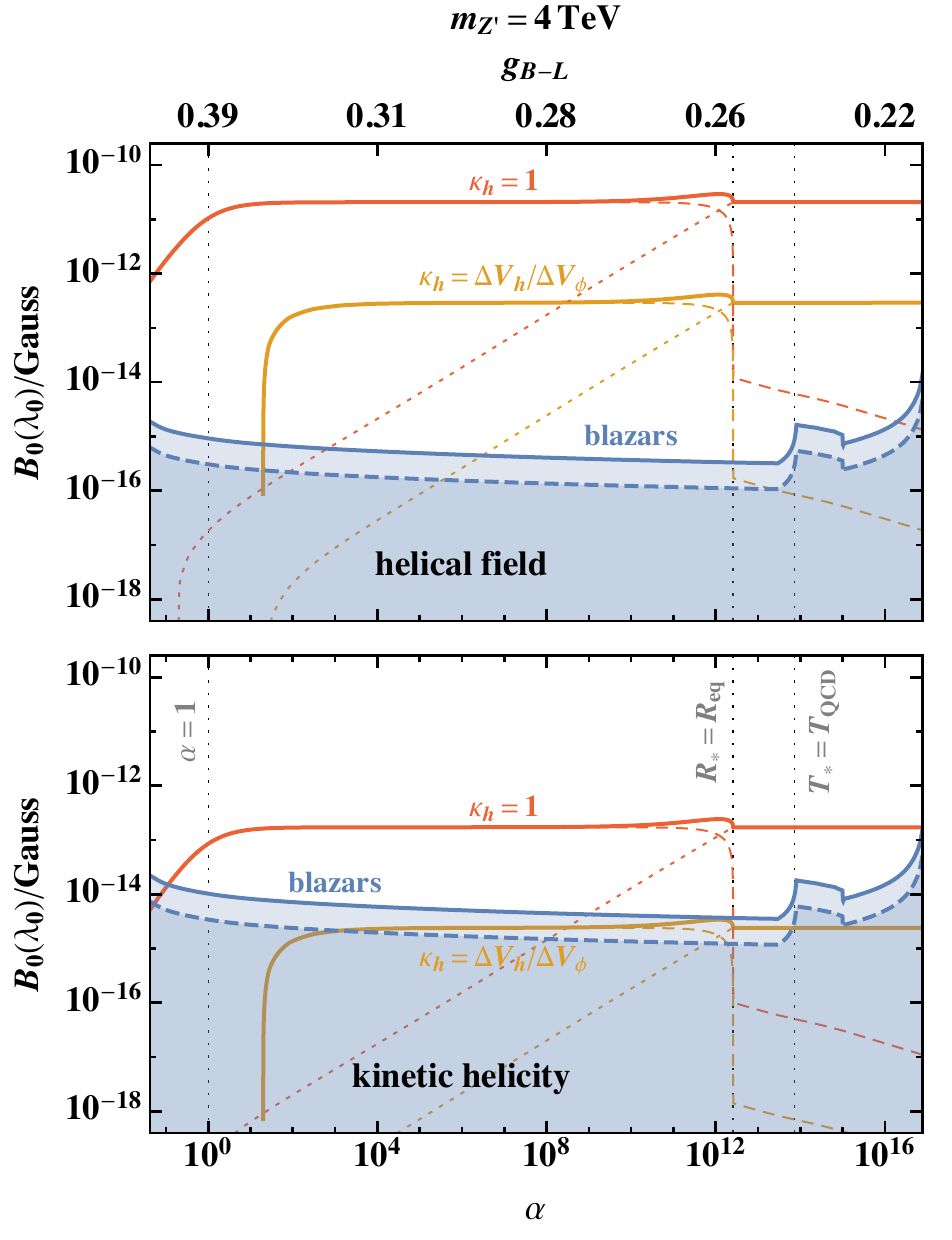} \hspace{2mm}
    \includegraphics[height=9.cm]{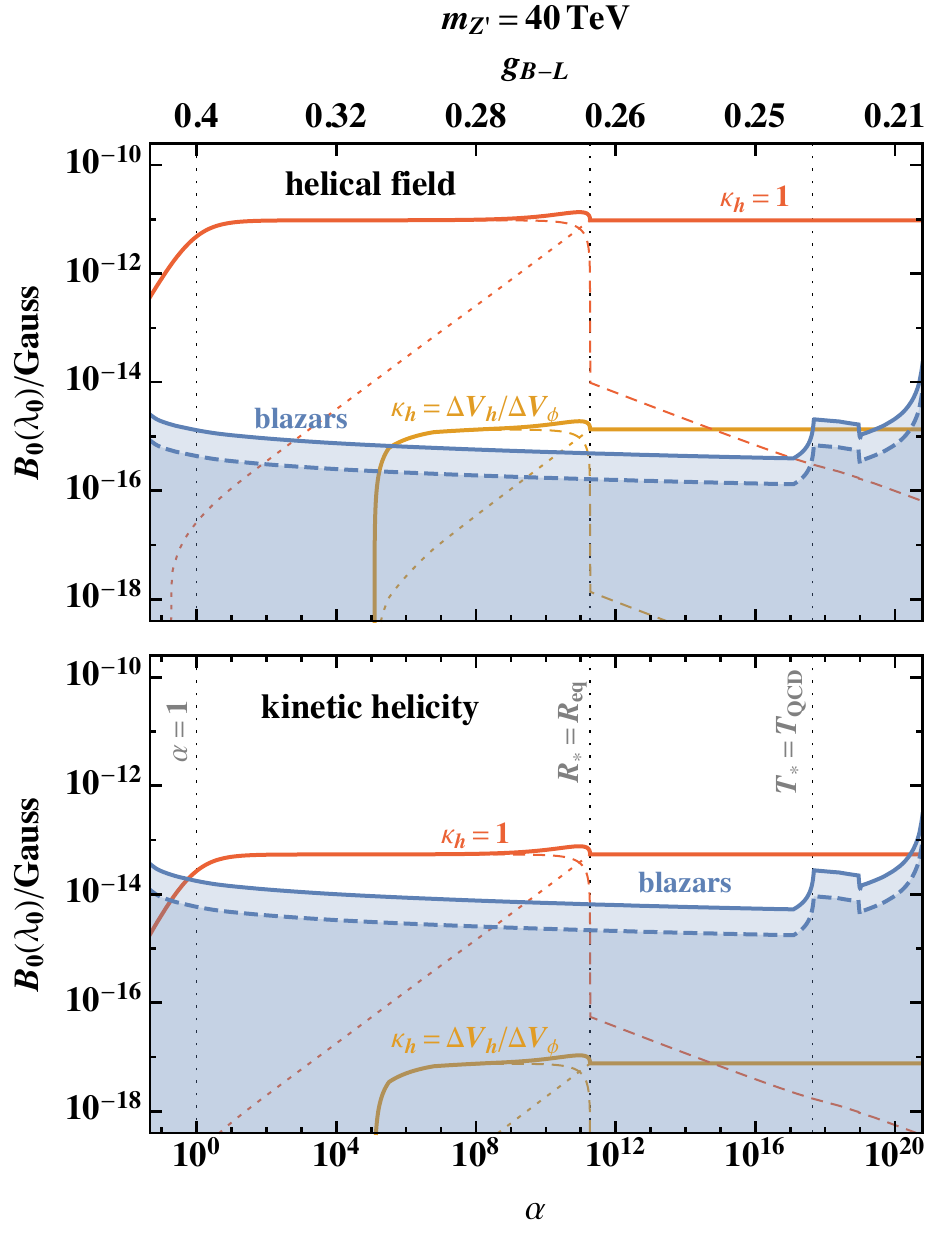}
    \caption{The red and yellow lines show the magnetic field strength $B_0$ at the coherence scale $\lambda_0$ with and without the $\Delta V_h/\Delta V_\varphi$ suppression. The short and long dashed lines show separately the contributions from bubble collisions and turbulence in the plasma. The upper and lower panels correspond respectively to evolution assuming a helical magnetic field or kinetic helicity, as shown in Fig.~\ref{fig:Bplotcsi}.}
    \label{fig:B0plotcsi}
\end{figure}

\begin{figure}
    \centering
    \includegraphics[height=9.cm]{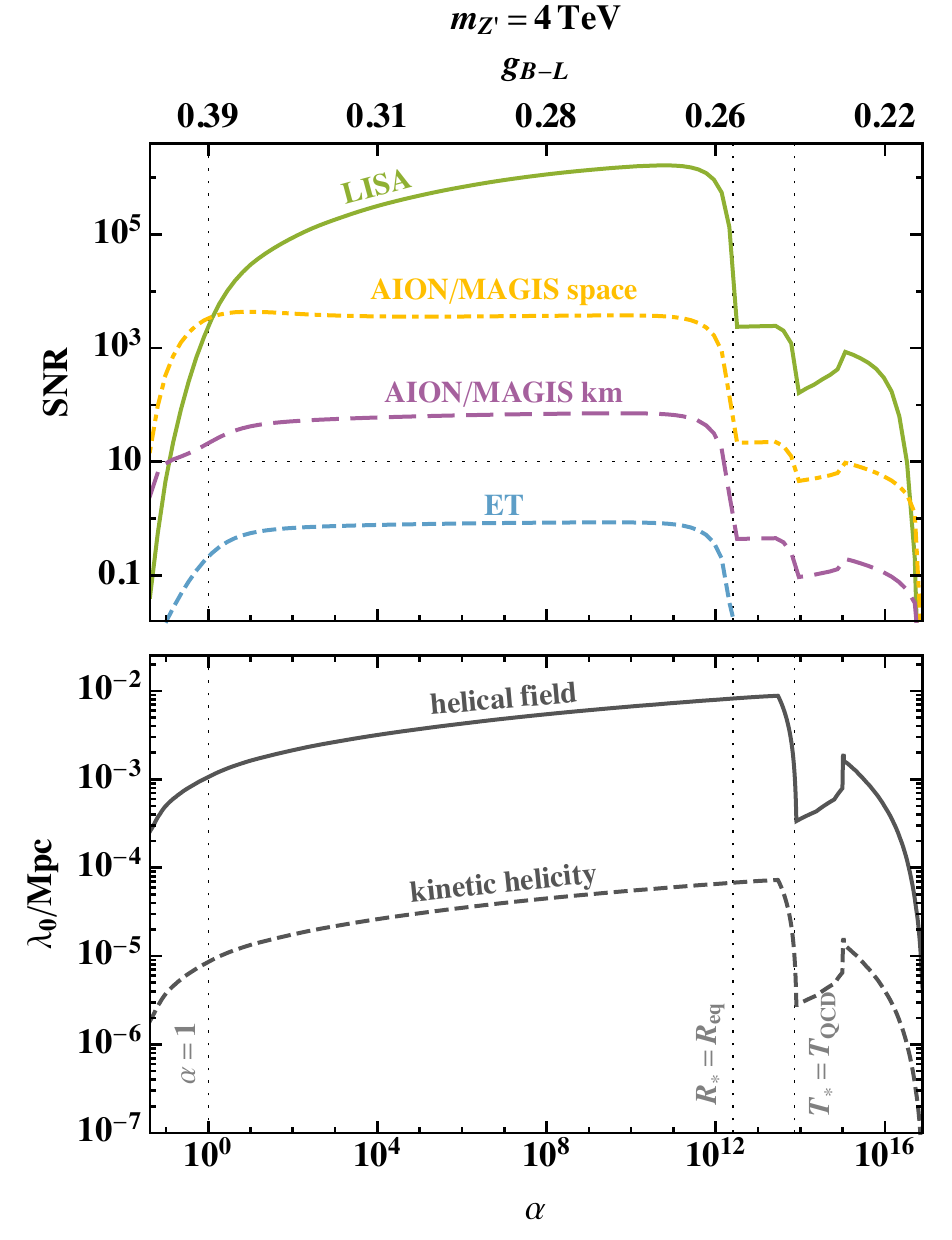} \hspace{2mm}
    \includegraphics[height=9.cm]{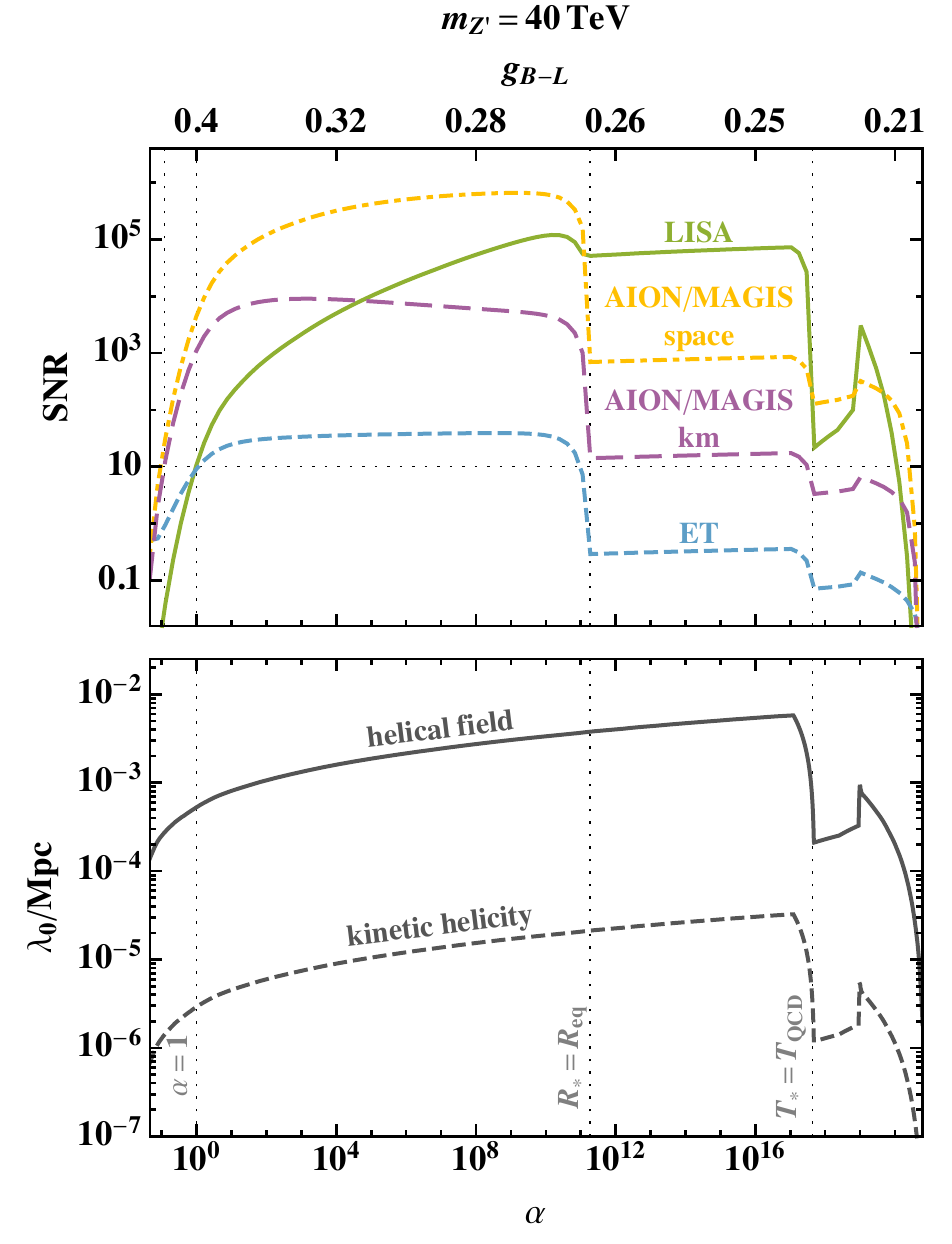}
    \caption{The lower panels show the coherence scale of the magnetic field produced in the phase transition for the two different evolutions shown in Fig.~\ref{fig:Bplotcsi}. The show the SNR of the GW signal from the phase transition for different future observatories.}
    \label{fig:lamnda0plotcsi}
\end{figure}

In the following we show results for $m_{Z'} = 4\,{\rm TeV}$ and $m_{Z'} = 40\,{\rm TeV}$, which correspond roughly to the masses currently being probed by the LHC and potentially observable at a 100\,TeV circular collider, respectively. Using the latest constraints from the ATLAS collaboration \cite{ATLAS:2019vcr}, we find that the 95\% confidence level constraints on the gauge coupling $g_{B-L}$ are \{0.11, 0.40, 0.83\} for $m_{Z'}=$\{3\,TeV, 4\,TeV, 5\,TeV\} respectively.

The production of magnetic fields in this model is somewhat less trivial than in SM extensions that modify the Higgs potential to feature a first-order phase transition. The reason is that most of the energy of the transition is carried by bubbles of the new scalar field $\varphi$, or is transferred to the dark sector $Z'$ plasma, neither of which contribute to magnetic fields directly.

If the transition proceeds at a sufficiently low temperature, $T \lesssim 140$\,GeV, a small fraction of the bubble wall energy is immediately stored in the Higgs field bubble, which follows the $\varphi$ bubble due to the portal coupling giving Higgs a negative mass term inside the new scalar bubble. This fraction is proportional to $\left(V(v_\varphi,0)-V(0,0)\right)/\left(V(v_\varphi,0)-V(v_\varphi,v_h)\right) \equiv \Delta V_h/\Delta V_\varphi$, which is small due to the large ratio between the vacuum expectation values of the fields. Provided that kinetic mixing between the U$(1)_Y$ and U$(1)_{B-L}$ gauge fields is present, a part of the energy deposited into the dark plasma is expected to be transferred to the SM fields. However, in Ref.~\cite{Kamada:2018kyi} it was shown that the magnetic field transfer from a dark U$(1)$ to the visible U$(1)$ through kinetic mixing is not efficient, and the transferred magnetic field is weaker than that generated directly.

In the same way as in the SM+$H^6$ model discussed in the previous subsection, we show the results for the magnetic field spectrum in Figs.~\ref{fig:Bplotcsi}, \ref{fig:B0plotcsi} and \ref{fig:lamnda0plotcsi}. 
The cases with $\kappa_h = \Delta V_h/\Delta V_\varphi$ show the magnetic field produced by the Higgs field bubbles. For comparison, we show also the overly optimistic case with no suppression, $\kappa_h=1$, that would correspond to the case where the transfer of the magnetic field from the dark U$(1)$ to the visible one would be very efficient.

We see in Fig.~\ref{fig:Bplotcsi} that both bubble collisions and turbulence may yield a peak above the blazar lower limit if the primordial field was seeded by magnetic helicity. The total strength of the magnetic field is shown as a function of the phase transition strength $\alpha$ in Fig.~\ref{fig:B0plotcsi}. We see that the total strength of the magnetic field for $\alpha\gg 1$ is roughly independent of the strength of the transition, with bubble collisions making the dominant contribution when the average radius of the bubbles at percolation is $R < R_{\rm eq}$ (corresponding to $\alpha \lesssim 10^{12}$), and turbulence dominating at larger $\alpha$. This is because to the right of the $R=R_{\rm eq}$ line the bubble walls do not reach terminal velocity before they collide, so most of the released vacuum energy is carried by the bubble wall, which is the dominant source for both the magnetic field and the GWs shown in the upper panels of Fig.~\ref{fig:lamnda0plotcsi}, whereas to the left of that line it is given by the plasma-related sources. The phase transition finishes only after the QCD phase transition for values of $\alpha$ to the right of the $T=T_{\rm QCD}$ line.

In the lower panels of Fig.~\ref{fig:lamnda0plotcsi} we see that, as in the SM+$H^6$ model, the coherence scale of the peak of the magnetic field is below $0.01$\,Mpc in all the cases studied. The upper panels of Fig.~\ref{fig:lamnda0plotcsi} show the SNR of the GW signal from the phase transition for different future observatories. Comparing the upper and lower panels of Fig.~\ref{fig:B0plotcsi}, we see that, taking into account the $\kappa_h = \Delta V_h/\Delta V_\varphi$ suppression factor, the magnetic field could lie well above the blazar lower bound if it was seeded by magnetic field helicity, but not if it was seeded by kinetic helicity.

After the phase transition the plasma is reheated to a high temperature where the electroweak symmetry is restored. Then, as the plasma cools, the electroweak symmetry becomes again broken, but this time the transition is a smooth crossover. The results of  Ref.~\cite{Kamada:2016cnb} indicate that in the presence of a helical magnetic field that could explain the blazar observations an overabundance of baryons is produced in the latter transition, which would place constraints on the helical field case shown in the upper panels of Fig.~\ref{fig:B0plotcsi}. The impact depends on the evolution of the weak mixing angle during the crossover, which has considerable uncertainty. However, it seems that a transition faster than in the SM case~\cite{DOnofrio:2015gop} is required in order to avoid the baryon overproduction.

\section{Conclusions}
\label{sec:conx}

In this work we have revisited the possibility of generating magnetic fields during a first-order phase transition in the early Universe.  Since this phase transition is second-order in the SM, we have considered two well motivated beyond the standard model extensions.  First we considered a dimension-6 extension of the Higgs potential that leads directly to a first-order phase transition.  We showed that, in such a model, the strong coupling between the Higgs and the rest of the SM would result in friction that would slow the motion of bubble walls, suppressing the subsequent generation of magnetic fields.  We also showed, however, that turbulence created by vorticity in gauge fields in the plasma could create magnetic fields with interesting magnitudes.  We also calculated the corresponding gravitational wave signal associated with such phase transitions. 

The second model we considered was a minimal extension of the SM with an additional scalar $\varphi$ associated with a $B-L$ gauge symmetry. In this situation, the field $\varphi$ can undergo a strong first-order phase transition where most of the energy goes into the bubble wall rather than the plasma.  A part of this energy is later transmitted to the SM fields via a portal coupling, potentially leading to large magnetic fields. Again we have calculated the gravitational wave signal for this model.  The strength of the magnetic field goes down as the mass of the Z' boson increases, and the existing limits on $m_{Z'}$ and $g_{B-L}$ from the LHC leave open a region where it is possible to explain the observed magnetic fields. However, this window could be further constrained by baryon overproduction.

In case of the $U(1)_{B-L}$ model, whenever we are able to produce magnetic fields with sufficient magnitude to explain the {\it Fermi} data, we produce enough GWs to be detected by one of the future experiments.
This is not the case in the $SM+H^6$ model, where GW experiments will be able to probe all of the relevant parameter space only if the magnetic field is seeded by kinetic helicity,
whereas a fully helical magnetic field could be produced without a corresponding GW signal 
being within reach. In this model, however, the HL-LHC will be able to probe all of the relevant parameter space. 
These conclusions suggest that, generically, the production of strong intergalactic magnetic fields through a phase transition in the early universe may lead to other observable signals either in future GW experiments or at a future collider.

The renewed interest in strong first-order phase transitions due to recent developments in GW detection shows no sign of diminishing.  When considering such phase transitions and their effects upon the SM gauge fields, it is natural to wonder if magnetic fields may have such a dramatic origin.  This paper shows that this is indeed possible under appropriate conditions.

\section*{Acknowledgements}

We are grateful for conversations with Kostas Dimopoulos.  The work of JE, MF, ML and VV was supported by the United Kingdom STFC Grant
ST/P000258/1. Also, JE received support from the Estonian Research Council via a
Mobilitas Pluss grant, ML was partly supported by the Polish National Science Center grant 2018/31/D/ST2/02048, and MF and AW were funded by the European Research Council under the European Union's Horizon 2020 programme (ERC Grant Agreement no.648680 DARKHORIZONS).

\bibliographystyle{JHEP}
\bibliography{Mag}

\providecommand{\href}[2]{#2}\begingroup\raggedright\begin{thebibliography}{10}

\bibitem{Grasso:2000wj}
D.~Grasso and H.~R. Rubinstein, \emph{{Magnetic fields in the early universe}},
  \href{http://dx.doi.org/10.1016/S0370-1573(00)00110-1}{\emph{Phys. Rept.}
  {\bf 348} (2001) 163--266},
  [\href{http://arxiv.org/abs/astro-ph/0009061}{{\tt astro-ph/0009061}}].

\bibitem{Kawasaki:2012va}
M.~Kawasaki and M.~Kusakabe, \emph{{Updated constraint on a primordial magnetic
  field during big bang nucleosynthesis and a formulation of field effects}},
  \href{http://dx.doi.org/10.1103/PhysRevD.86.063003}{\emph{Phys. Rev.} {\bf
  D86} (2012) 063003}, [\href{http://arxiv.org/abs/1204.6164}{{\tt
  1204.6164}}].

\bibitem{Seshadri:2009sy}
T.~R. Seshadri and K.~Subramanian, \emph{{CMB bispectrum from primordial
  magnetic fields on large angular scales}},
  \href{http://dx.doi.org/10.1103/PhysRevLett.103.081303}{\emph{Phys. Rev.
  Lett.} {\bf 103} (2009) 081303}, [\href{http://arxiv.org/abs/0902.4066}{{\tt
  0902.4066}}].

\bibitem{Ade:2015cva}
{\scshape Planck} collaboration, P.~A.~R. Ade et~al., \emph{{Planck 2015
  results. XIX. Constraints on primordial magnetic fields}},
  \href{http://dx.doi.org/10.1051/0004-6361/201525821}{\emph{Astron.
  Astrophys.} {\bf 594} (2016) A19},
  [\href{http://arxiv.org/abs/1502.01594}{{\tt 1502.01594}}].

\bibitem{Jedamzik:1996wp}
K.~Jedamzik, V.~Katalinic and A.~V. Olinto, \emph{{Damping of cosmic magnetic
  fields}}, \href{http://dx.doi.org/10.1103/PhysRevD.57.3264}{\emph{Phys. Rev.}
  {\bf D57} (1998) 3264--3284},
  [\href{http://arxiv.org/abs/astro-ph/9606080}{{\tt astro-ph/9606080}}].

\bibitem{Jedamzik:1999bm}
K.~Jedamzik, V.~Katalinic and A.~V. Olinto, \emph{{A Limit on primordial small
  scale magnetic fields from CMB distortions}},
  \href{http://dx.doi.org/10.1103/PhysRevLett.85.700}{\emph{Phys. Rev. Lett.}
  {\bf 85} (2000) 700--703}, [\href{http://arxiv.org/abs/astro-ph/9911100}{{\tt
  astro-ph/9911100}}].

\bibitem{Barrow:1997mj}
J.~D. Barrow, P.~G. Ferreira and J.~Silk, \emph{{Constraints on a primordial
  magnetic field}},
  \href{http://dx.doi.org/10.1103/PhysRevLett.78.3610}{\emph{Phys. Rev. Lett.}
  {\bf 78} (1997) 3610--3613},
  [\href{http://arxiv.org/abs/astro-ph/9701063}{{\tt astro-ph/9701063}}].

\bibitem{Durrer:1999bk}
R.~Durrer, P.~G. Ferreira and T.~Kahniashvili, \emph{{Tensor microwave
  anisotropies from a stochastic magnetic field}},
  \href{http://dx.doi.org/10.1103/PhysRevD.61.043001}{\emph{Phys. Rev.} {\bf
  D61} (2000) 043001}, [\href{http://arxiv.org/abs/astro-ph/9911040}{{\tt
  astro-ph/9911040}}].

\bibitem{Yamazaki:2012pg}
D.~G. Yamazaki, T.~Kajino, G.~J. Mathew and K.~Ichiki, \emph{{The Search for a
  Primordial Magnetic Field}},
  \href{http://dx.doi.org/10.1016/j.physrep.2012.02.005}{\emph{Phys. Rept.}
  {\bf 517} (2012) 141--167}, [\href{http://arxiv.org/abs/1204.3669}{{\tt
  1204.3669}}].

\bibitem{Trivedi:2010gi}
P.~Trivedi, K.~Subramanian and T.~R. Seshadri, \emph{{Primordial Magnetic Field
  Limits from Cosmic Microwave Background Bispectrum of Magnetic Passive Scalar
  Modes}}, \href{http://dx.doi.org/10.1103/PhysRevD.82.123006}{\emph{Phys.
  Rev.} {\bf D82} (2010) 123006}, [\href{http://arxiv.org/abs/1009.2724}{{\tt
  1009.2724}}].

\bibitem{Tavecchio:2010mk}
F.~Tavecchio, G.~Ghisellini, L.~Foschini, G.~Bonnoli, G.~Ghirlanda and
  P.~Coppi, \emph{{The intergalactic magnetic field constrained by Fermi/LAT
  observations of the TeV blazar 1ES 0229+200}},
  \href{http://dx.doi.org/10.1111/j.1745-3933.2010.00884.x}{\emph{Mon. Not.
  Roy. Astron. Soc.} {\bf 406} (2010) L70--L74},
  [\href{http://arxiv.org/abs/1004.1329}{{\tt 1004.1329}}].

\bibitem{Ando:2010rb}
S.~Ando and A.~Kusenko, \emph{{Evidence for Gamma-Ray Halos Around Active
  Galactic Nuclei and the First Measurement of Intergalactic Magnetic Fields}},
  \href{http://dx.doi.org/10.1088/2041-8205/722/1/L39}{\emph{Astrophys. J.}
  {\bf 722} (2010) L39}, [\href{http://arxiv.org/abs/1005.1924}{{\tt
  1005.1924}}].

\bibitem{Neronov:1900zz}
A.~Neronov and I.~Vovk, \emph{{Evidence for strong extragalactic magnetic
  fields from Fermi observations of TeV blazars}},
  \href{http://dx.doi.org/10.1126/science.1184192}{\emph{Science} {\bf 328}
  (2010) 73--75}, [\href{http://arxiv.org/abs/1006.3504}{{\tt 1006.3504}}].

\bibitem{Essey:2010nd}
W.~Essey, S.~Ando and A.~Kusenko, \emph{{Determination of intergalactic
  magnetic fields from gamma ray data}},
  \href{http://dx.doi.org/10.1016/j.astropartphys.2011.06.010}{\emph{Astropart.
  Phys.} {\bf 35} (2011) 135--139}, [\href{http://arxiv.org/abs/1012.5313}{{\tt
  1012.5313}}].

\bibitem{Chen:2014rsa}
W.~Chen, J.~H. Buckley and F.~Ferrer, \emph{{Search for GeV γ -Ray Pair Halos
  Around Low Redshift Blazars}},
  \href{http://dx.doi.org/10.1103/PhysRevLett.115.211103}{\emph{Phys. Rev.
  Lett.} {\bf 115} (2015) 211103}, [\href{http://arxiv.org/abs/1410.7717}{{\tt
  1410.7717}}].

\bibitem{Biteau:2018tmv}
{\scshape Fermi-LAT} collaboration, M.~Ackermann et~al., \emph{{The Search for
  Spatial Extension in High-latitude Sources Detected by the $Fermi$ Large Area
  Telescope}},
  \href{http://dx.doi.org/10.3847/1538-4365/aacdf7}{\emph{Astrophys. J. Suppl.}
  {\bf 237} (2018) 32}, [\href{http://arxiv.org/abs/1804.08035}{{\tt
  1804.08035}}].

\bibitem{1950ZNatA...5..237S}
A.~{Schl{\"u}ter} and L.~{Biermann}, \emph{{Interstellare Magnetfelder}},
  \href{http://dx.doi.org/10.1515/zna-1950-0501}{\emph{Zeitschrift
  Naturforschung Teil A} {\bf 5} (May, 1950) 237--251}.

\bibitem{Kulsrud:1996km}
R.~M. Kulsrud, R.~Cen, J.~P. Ostriker and D.~Ryu, \emph{{The Protogalactic
  origin for cosmic magnetic fields}},
  \href{http://dx.doi.org/10.1086/303987}{\emph{Astrophys. J.} {\bf 480} (1997)
  481}, [\href{http://arxiv.org/abs/astro-ph/9607141}{{\tt astro-ph/9607141}}].

\bibitem{Gnedin:2000ax}
N.~Y. Gnedin, A.~Ferrara and E.~G. Zweibel, \emph{{Generation of the primordial
  magnetic fields during cosmological reionization}},
  \href{http://dx.doi.org/10.1086/309272}{\emph{Astrophys. J.} {\bf 539} (2000)
  505--516}, [\href{http://arxiv.org/abs/astro-ph/0001066}{{\tt
  astro-ph/0001066}}].

\bibitem{1984JFM...144....1Z}
Y.~B. {Zel'dovich}, A.~A. {Ruzmaikin}, S.~A. {Molchanov} and D.~D. {Sokolov},
  \emph{{Kinematic dynamo problem in a linear velocity field}},
  \href{http://dx.doi.org/10.1017/S0022112084001488}{\emph{Journal of Fluid
  Mechanics} {\bf 144} (Jan, 1984) 1--11}.

\bibitem{Shukurov:2005pf}
A.~Shukurov, D.~Sokoloff, K.~Subramanian and A.~Brandenburg, \emph{{Galactic
  dynamo and helicity losses through fountain flow}},
  \href{http://dx.doi.org/10.1051/0004-6361:200600011}{\emph{Astron.
  Astrophys.} {\bf 448} (2006) L33--L36},
  [\href{http://arxiv.org/abs/astro-ph/0512592}{{\tt astro-ph/0512592}}].

\bibitem{Kulsrud:2007an}
R.~M. Kulsrud and E.~G. Zweibel, \emph{{The Origin of Astrophysical Magnetic
  Fields}}, \href{http://dx.doi.org/10.1088/0034-4885/71/4/046901}{\emph{Rept.
  Prog. Phys.} {\bf 71} (2008) 0046091},
  [\href{http://arxiv.org/abs/0707.2783}{{\tt 0707.2783}}].

\bibitem{Turner:1987bw}
M.~S. Turner and L.~M. Widrow, \emph{{Inflation Produced, Large Scale Magnetic
  Fields}}, \href{http://dx.doi.org/10.1103/PhysRevD.37.2743}{\emph{Phys. Rev.}
  {\bf D37} (1988) 2743}.

\bibitem{Ratra:1991bn}
B.~Ratra, \emph{{Cosmological 'seed' magnetic field from inflation}},
  \href{http://dx.doi.org/10.1086/186384}{\emph{Astrophys. J.} {\bf 391} (1992)
  L1--L4}.

\bibitem{Martin:2007ue}
J.~Martin and J.~Yokoyama, \emph{{Generation of Large-Scale Magnetic Fields in
  Single-Field Inflation}},
  \href{http://dx.doi.org/10.1088/1475-7516/2008/01/025}{\emph{JCAP} {\bf 0801}
  (2008) 025}, [\href{http://arxiv.org/abs/0711.4307}{{\tt 0711.4307}}].

\bibitem{Kobayashi:2014sga}
T.~Kobayashi, \emph{{Primordial Magnetic Fields from the Post-Inflationary
  Universe}},
  \href{http://dx.doi.org/10.1088/1475-7516/2014/05/040}{\emph{JCAP} {\bf 1405}
  (2014) 040}, [\href{http://arxiv.org/abs/1403.5168}{{\tt 1403.5168}}].

\bibitem{Vachaspati:1991nm}
T.~Vachaspati, \emph{{Magnetic fields from cosmological phase transitions}},
  \href{http://dx.doi.org/10.1016/0370-2693(91)90051-Q}{\emph{Phys. Lett.} {\bf
  B265} (1991) 258--261}.

\bibitem{Sigl:1996dm}
G.~Sigl, A.~V. Olinto and K.~Jedamzik, \emph{{Primordial magnetic fields from
  cosmological first order phase transitions}},
  \href{http://dx.doi.org/10.1103/PhysRevD.55.4582}{\emph{Phys. Rev.} {\bf D55}
  (1997) 4582--4590}, [\href{http://arxiv.org/abs/astro-ph/9610201}{{\tt
  astro-ph/9610201}}].

\bibitem{Tevzadze:2012kk}
A.~G. Tevzadze, L.~Kisslinger, A.~Brandenburg and T.~Kahniashvili,
  \emph{{Magnetic Fields from QCD Phase Transitions}},
  \href{http://dx.doi.org/10.1088/0004-637X/759/1/54}{\emph{Astrophys. J.} {\bf
  759} (2012) 54}, [\href{http://arxiv.org/abs/1207.0751}{{\tt 1207.0751}}].

\bibitem{Espinosa:2011ax}
J.~R. Espinosa, T.~Konstandin and F.~Riva, \emph{{Strong Electroweak Phase
  Transitions in the Standard Model with a Singlet}},
  \href{http://dx.doi.org/10.1016/j.nuclphysb.2011.09.010}{\emph{Nucl. Phys.}
  {\bf B854} (2012) 592--630}, [\href{http://arxiv.org/abs/1107.5441}{{\tt
  1107.5441}}].

\bibitem{Kuzmin:1985mm}
V.~A. Kuzmin, V.~A. Rubakov and M.~E. Shaposhnikov, \emph{{On the Anomalous
  Electroweak Baryon Number Nonconservation in the Early Universe}},
  \href{http://dx.doi.org/10.1016/0370-2693(85)91028-7}{\emph{Phys. Lett.} {\bf
  155B} (1985) 36}.

\bibitem{Fujita:2016igl}
T.~Fujita and K.~Kamada, \emph{{Large-scale magnetic fields can explain the
  baryon asymmetry of the Universe}},
  \href{http://dx.doi.org/10.1103/PhysRevD.93.083520}{\emph{Phys. Rev.} {\bf
  D93} (2016) 083520}, [\href{http://arxiv.org/abs/1602.02109}{{\tt
  1602.02109}}].

\bibitem{Kamada:2016eeb}
K.~Kamada and A.~J. Long, \emph{{Baryogenesis from decaying magnetic
  helicity}}, \href{http://dx.doi.org/10.1103/PhysRevD.94.063501}{\emph{Phys.
  Rev.} {\bf D94} (2016) 063501}, [\href{http://arxiv.org/abs/1606.08891}{{\tt
  1606.08891}}].

\bibitem{Kamada:2016cnb}
K.~Kamada and A.~J. Long, \emph{{Evolution of the Baryon Asymmetry through the
  Electroweak Crossover in the Presence of a Helical Magnetic Field}},
  \href{http://dx.doi.org/10.1103/PhysRevD.94.123509}{\emph{Phys. Rev.} {\bf
  D94} (2016) 123509}, [\href{http://arxiv.org/abs/1610.03074}{{\tt
  1610.03074}}].

\bibitem{Witten:1984rs}
E.~Witten, \emph{{Cosmic Separation of Phases}},
  \href{http://dx.doi.org/10.1103/PhysRevD.30.272}{\emph{Phys. Rev.} {\bf D30}
  (1984) 272--285}.

\bibitem{Ellis:2018mja}
J.~Ellis, M.~Lewicki and J.~M. No, \emph{{On the Maximal Strength of a
  First-Order Electroweak Phase Transition and its Gravitational Wave Signal}},
  \href{http://dx.doi.org/10.1088/1475-7516/2019/04/003}{\emph{JCAP} {\bf 1904}
  (2019) 003}, [\href{http://arxiv.org/abs/1809.08242}{{\tt 1809.08242}}].

\bibitem{Ellis:2019oqb}
J.~Ellis, M.~Lewicki, J.~M. No and V.~Vaskonen, \emph{{Gravitational wave
  energy budget in strongly supercooled phase transitions}},
  \href{http://dx.doi.org/10.1088/1475-7516/2019/06/024}{\emph{JCAP} {\bf 1906}
  (2019) 024}, [\href{http://arxiv.org/abs/1903.09642}{{\tt 1903.09642}}].

\bibitem{Cornwall:1997ms}
J.~M. Cornwall, \emph{{Speculations on primordial magnetic helicity}},
  \href{http://dx.doi.org/10.1103/PhysRevD.56.6146}{\emph{Phys. Rev.} {\bf D56}
  (1997) 6146--6154}, [\href{http://arxiv.org/abs/hep-th/9704022}{{\tt
  hep-th/9704022}}].

\bibitem{Giovannini:1997gp}
M.~Giovannini and M.~E. Shaposhnikov, \emph{{Primordial magnetic fields,
  anomalous isocurvature fluctuations and big bang nucleosynthesis}},
  \href{http://dx.doi.org/10.1103/PhysRevLett.80.22}{\emph{Phys. Rev. Lett.}
  {\bf 80} (1998) 22--25}, [\href{http://arxiv.org/abs/hep-ph/9708303}{{\tt
  hep-ph/9708303}}].

\bibitem{1996PhRvE..53.1283S}
N.~Seehafer, \emph{Nature of the \ensuremath{\alpha} effect in
  magnetohydrodynamics},
  \href{http://dx.doi.org/10.1103/PhysRevE.53.1283}{\emph{Phys. Rev. E} {\bf
  53} (Jan, 1996) 1283--1286}.

\bibitem{1999PhRvL..83.3198J}
H.~Ji, \emph{{Turbulent dynamos and magnetic helicity}},
  \href{http://dx.doi.org/10.1103/PhysRevLett.83.3198}{\emph{Phys. Rev. Lett.}
  {\bf 83} (1999) 3198--3201},
  [\href{http://arxiv.org/abs/astro-ph/0102321}{{\tt astro-ph/0102321}}].

\bibitem{1981PhRvL..47.1060M}
A.~Pouquet, M.~Meneguzzi and U.~Frisch, \emph{{Growth of correlations in
  magnetohydrodynamic turbulence}},
  \href{http://dx.doi.org/10.1103/PhysRevA.33.4266}{\emph{Phys. Rev.} {\bf A33}
  (1986) 4266--4276}.

\bibitem{Zhang:2019vsb}
Y.~Zhang, T.~Vachaspati and F.~Ferrer, \emph{{Magnetic field production at a
  first-order electroweak phase transition}},
  \href{http://arxiv.org/abs/1902.02751}{{\tt 1902.02751}}.

\bibitem{Espinosa:2010hh}
J.~R. Espinosa, T.~Konstandin, J.~M. No and G.~Servant, \emph{{Energy Budget of
  Cosmological First-order Phase Transitions}},
  \href{http://dx.doi.org/10.1088/1475-7516/2010/06/028}{\emph{JCAP} {\bf 1006}
  (2010) 028}, [\href{http://arxiv.org/abs/1004.4187}{{\tt 1004.4187}}].

\bibitem{Caprini:2015zlo}
C.~Caprini et~al., \emph{{Science with the space-based interferometer eLISA.
  II: Gravitational waves from cosmological phase transitions}},
  \href{http://dx.doi.org/10.1088/1475-7516/2016/04/001}{\emph{JCAP} {\bf 1604}
  (2016) 001}, [\href{http://arxiv.org/abs/1512.06239}{{\tt 1512.06239}}].

\bibitem{Kahniashvili:2009qi}
T.~Kahniashvili, A.~G. Tevzadze and B.~Ratra, \emph{{Phase Transition Generated
  Cosmological Magnetic Field at Large Scales}},
  \href{http://dx.doi.org/10.1088/0004-637X/726/2/78}{\emph{Astrophys. J.} {\bf
  726} (2011) 78}, [\href{http://arxiv.org/abs/0907.0197}{{\tt 0907.0197}}].

\bibitem{Durrer:2013pga}
R.~Durrer and A.~Neronov, \emph{{Cosmological Magnetic Fields: Their
  Generation, Evolution and Observation}},
  \href{http://dx.doi.org/10.1007/s00159-013-0062-7}{\emph{Astron. Astrophys.
  Rev.} {\bf 21} (2013) 62}, [\href{http://arxiv.org/abs/1303.7121}{{\tt
  1303.7121}}].

\bibitem{Brandenburg:2017neh}
A.~Brandenburg, T.~Kahniashvili, S.~Mandal, A.~R. Pol, A.~G. Tevzadze and
  T.~Vachaspati, \emph{{Evolution of hydromagnetic turbulence from the
  electroweak phase transition}},
  \href{http://dx.doi.org/10.1103/PhysRevD.96.123528}{\emph{Phys. Rev.} {\bf
  D96} (2017) 123528}, [\href{http://arxiv.org/abs/1711.03804}{{\tt
  1711.03804}}].

\bibitem{Durrer:2003ja}
R.~Durrer and C.~Caprini, \emph{{Primordial magnetic fields and causality}},
  \href{http://dx.doi.org/10.1088/1475-7516/2003/11/010}{\emph{JCAP} {\bf 0311}
  (2003) 010}, [\href{http://arxiv.org/abs/astro-ph/0305059}{{\tt
  astro-ph/0305059}}].

\bibitem{Kolmogorov:1941}
A.~N. Kolmogorov, \emph{The local structure of turbulence in incompressible
  viscous fluid for very large reynolds numbers}, {\emph{Proceedings:
  Mathematical and Physical Sciences} {\bf 434} (1991) 9--13}.

\bibitem{Landau1987Fluid}
L.~D. Landau and E.~M. Lifschitz, \emph{{Fluid Mechanics}}.
\newblock 1987.

\bibitem{Biskamp:1999}
D.~Biskamp and W.-C. M\"uller, \emph{Decay laws for three-dimensional
  magnetohydrodynamic turbulence},
  \href{http://dx.doi.org/10.1103/PhysRevLett.83.2195}{\emph{Phys. Rev. Lett.}
  {\bf 83} (Sep, 1999) 2195--2198}.

\bibitem{biskamp_2003}
D.~Biskamp, \emph{Magnetohydrodynamic Turbulence}.
\newblock Cambridge University Press, 2003,
  \href{http://dx.doi.org/10.1017/CBO9780511535222}{10.1017/CBO9780511535222}.

\bibitem{Vachaspati:2001nb}
T.~Vachaspati, \emph{{Estimate of the primordial magnetic field helicity}},
  \href{http://dx.doi.org/10.1103/PhysRevLett.87.251302}{\emph{Phys. Rev.
  Lett.} {\bf 87} (2001) 251302},
  [\href{http://arxiv.org/abs/astro-ph/0101261}{{\tt astro-ph/0101261}}].

\bibitem{Brandenburg:2017rnt}
A.~Brandenburg, T.~Kahniashvili, S.~Mandal, A.~R. Pol, A.~G. Tevzadze and
  T.~Vachaspati, \emph{{The dynamo effect in decaying helical turbulence}},
  \href{http://dx.doi.org/10.1103/PhysRevFluids.4.024608}{\emph{Phys. Rev.
  Fluids.} {\bf 4} (2019) 024608}, [\href{http://arxiv.org/abs/1710.01628}{{\tt
  1710.01628}}].

\bibitem{Dolag:2010ni}
K.~Dolag, M.~Kachelriess, S.~Ostapchenko and R.~Tomas, \emph{{Lower limit on
  the strength and filling factor of extragalactic magnetic fields}},
  \href{http://dx.doi.org/10.1088/2041-8205/727/1/L4}{\emph{Astrophys. J.} {\bf
  727} (2011) L4}, [\href{http://arxiv.org/abs/1009.1782}{{\tt 1009.1782}}].

\bibitem{Taylor:2011bn}
A.~M. Taylor, I.~Vovk and A.~Neronov, \emph{{Extragalactic magnetic fields
  constraints from simultaneous GeV-TeV observations of blazars}},
  \href{http://dx.doi.org/10.1051/0004-6361/201116441}{\emph{Astron.
  Astrophys.} {\bf 529} (2011) A144},
  [\href{http://arxiv.org/abs/1101.0932}{{\tt 1101.0932}}].

\bibitem{Dermer:2010mm}
C.~D. Dermer, M.~Cavadini, S.~Razzaque, J.~D. Finke, J.~Chiang and B.~Lott,
  \emph{{Time Delay of Cascade Radiation for TeV Blazars and the Measurement of
  the Intergalactic Magnetic Field}},
  \href{http://dx.doi.org/10.1088/2041-8205/733/2/L21}{\emph{Astrophys. J.}
  {\bf 733} (2011) L21}, [\href{http://arxiv.org/abs/1011.6660}{{\tt
  1011.6660}}].

\bibitem{Kahniashvili:2008pe}
T.~Kahniashvili, L.~Campanelli, G.~Gogoberidze, Y.~Maravin and B.~Ratra,
  \emph{{Gravitational Radiation from Primordial Helical Inverse Cascade MHD
  Turbulence}}, \href{http://dx.doi.org/10.1103/PhysRevD.78.123006,
  10.1103/PhysRevD.79.109901}{\emph{Phys. Rev.} {\bf D78} (2008) 123006},
  [\href{http://arxiv.org/abs/0809.1899}{{\tt 0809.1899}}].

\bibitem{Kisslinger:2015hua}
L.~Kisslinger and T.~Kahniashvili, \emph{{Polarized Gravitational Waves from
  Cosmological Phase Transitions}},
  \href{http://dx.doi.org/10.1103/PhysRevD.92.043006}{\emph{Phys. Rev.} {\bf
  D92} (2015) 043006}, [\href{http://arxiv.org/abs/1505.03680}{{\tt
  1505.03680}}].

\bibitem{Saga:2018ont}
S.~Saga, H.~Tashiro and S.~Yokoyama, \emph{{Limits on primordial magnetic
  fields from direct detection experiments of gravitational wave background}},
  \href{http://dx.doi.org/10.1103/PhysRevD.98.083518}{\emph{Phys. Rev.} {\bf
  D98} (2018) 083518}, [\href{http://arxiv.org/abs/1807.00561}{{\tt
  1807.00561}}].

\bibitem{Huang:2015izx}
F.~P. Huang, P.-H. Gu, P.-F. Yin, Z.-H. Yu and X.~Zhang, \emph{{Testing the
  electroweak phase transition and electroweak baryogenesis at the LHC and a
  circular electron-positron collider}},
  \href{http://dx.doi.org/10.1103/PhysRevD.93.103515}{\emph{Phys. Rev.} {\bf
  D93} (2016) 103515}, [\href{http://arxiv.org/abs/1511.03969}{{\tt
  1511.03969}}].

\bibitem{Artymowski:2016tme}
M.~Artymowski, M.~Lewicki and J.~D. Wells, \emph{{Gravitational wave and
  collider implications of electroweak baryogenesis aided by non-standard
  cosmology}}, \href{http://dx.doi.org/10.1007/JHEP03(2017)066}{\emph{JHEP}
  {\bf 03} (2017) 066}, [\href{http://arxiv.org/abs/1609.07143}{{\tt
  1609.07143}}].

\bibitem{Chala:2018ari}
M.~Chala, C.~Krause and G.~Nardini, \emph{{Signals of the electroweak phase
  transition at colliders and gravitational wave observatories}},
  \href{http://dx.doi.org/10.1007/JHEP07(2018)062}{\emph{JHEP} {\bf 07} (2018)
  062}, [\href{http://arxiv.org/abs/1802.02168}{{\tt 1802.02168}}].

\bibitem{Bartolo:2016ami}
N.~Bartolo et~al., \emph{{Science with the space-based interferometer LISA. IV:
  Probing inflation with gravitational waves}},
  \href{http://dx.doi.org/10.1088/1475-7516/2016/12/026}{\emph{JCAP} {\bf 1612}
  (2016) 026}, [\href{http://arxiv.org/abs/1610.06481}{{\tt 1610.06481}}].

\bibitem{Graham:2016plp}
P.~W. Graham, J.~M. Hogan, M.~A. Kasevich and S.~Rajendran, \emph{{Resonant
  mode for gravitational wave detectors based on atom interferometry}},
  \href{http://dx.doi.org/10.1103/PhysRevD.94.104022}{\emph{Phys. Rev.} {\bf
  D94} (2016) 104022}, [\href{http://arxiv.org/abs/1606.01860}{{\tt
  1606.01860}}].

\bibitem{Graham:2017pmn}
{\scshape MAGIS} collaboration, P.~W. Graham, J.~M. Hogan, M.~A. Kasevich,
  S.~Rajendran and R.~W. Romani, \emph{{Mid-band gravitational wave detection
  with precision atomic sensors}},  \href{http://arxiv.org/abs/1711.02225}{{\tt
  1711.02225}}.

\bibitem{AION:2018}
O.~Buchmuller, ``{The Atom Interferometer Observatory Network}.''
  \url{https://indico.cern.ch/event/760005/contributions/3152426/attachments/1735965/2807829/AION-Oxford-17102018.pptx.pdf},
  2018.

\bibitem{Bodeker:2004ws}
D.~Bodeker, L.~Fromme, S.~J. Huber and M.~Seniuch, \emph{{The Baryon asymmetry
  in the standard model with a low cut-off}},
  \href{http://dx.doi.org/10.1088/1126-6708/2005/02/026}{\emph{JHEP} {\bf 02}
  (2005) 026}, [\href{http://arxiv.org/abs/hep-ph/0412366}{{\tt
  hep-ph/0412366}}].

\bibitem{Delaunay:2007wb}
C.~Delaunay, C.~Grojean and J.~D. Wells, \emph{{Dynamics of Non-renormalizable
  Electroweak Symmetry Breaking}},
  \href{http://dx.doi.org/10.1088/1126-6708/2008/04/029}{\emph{JHEP} {\bf 04}
  (2008) 029}, [\href{http://arxiv.org/abs/0711.2511}{{\tt 0711.2511}}].

\bibitem{Curtin:2014jma}
D.~Curtin, P.~Meade and C.-T. Yu, \emph{{Testing Electroweak Baryogenesis with
  Future Colliders}},
  \href{http://dx.doi.org/10.1007/JHEP11(2014)127}{\emph{JHEP} {\bf 11} (2014)
  127}, [\href{http://arxiv.org/abs/1409.0005}{{\tt 1409.0005}}].

\bibitem{Ashoorioon:2009nf}
A.~Ashoorioon and T.~Konstandin, \emph{{Strong electroweak phase transitions
  without collider traces}},
  \href{http://dx.doi.org/10.1088/1126-6708/2009/07/086}{\emph{JHEP} {\bf 07}
  (2009) 086}, [\href{http://arxiv.org/abs/0904.0353}{{\tt 0904.0353}}].

\bibitem{Beniwal:2017eik}
A.~Beniwal, M.~Lewicki, J.~D. Wells, M.~White and A.~G. Williams,
  \emph{{Gravitational wave, collider and dark matter signals from a scalar
  singlet electroweak baryogenesis}},
  \href{http://dx.doi.org/10.1007/JHEP08(2017)108}{\emph{JHEP} {\bf 08} (2017)
  108}, [\href{http://arxiv.org/abs/1702.06124}{{\tt 1702.06124}}].

\bibitem{Iso:2009ss}
S.~Iso, N.~Okada and Y.~Orikasa, \emph{{Classically conformal $B^-$ L extended
  Standard Model}},
  \href{http://dx.doi.org/10.1016/j.physletb.2009.04.046}{\emph{Phys. Lett.}
  {\bf B676} (2009) 81--87}, [\href{http://arxiv.org/abs/0902.4050}{{\tt
  0902.4050}}].

\bibitem{Marzo:2018nov}
C.~Marzo, L.~Marzola and V.~Vaskonen, \emph{{Phase transition and vacuum
  stability in the classically conformal B-L model}},
  \href{http://arxiv.org/abs/1811.11169}{{\tt 1811.11169}}.

\bibitem{ATLAS:2019vcr}
{\scshape ATLAS} collaboration, \emph{{Search for high-mass dilepton resonances
  using $139\,\mathrm{fb}^{-1}$ of $pp$ collision data collected at
  $\sqrt{s}=13\,\mathrm{TeV}$ with the ATLAS detector}},
  {\emph{ATLAS-CONF-2019-001} (2019) }.

\bibitem{Kamada:2018kyi}
K.~Kamada, Y.~Tsai and T.~Vachaspati, \emph{{Magnetic Field Transfer From A
  Hidden Sector}},
  \href{http://dx.doi.org/10.1103/PhysRevD.98.043501}{\emph{Phys. Rev.} {\bf
  D98} (2018) 043501}, [\href{http://arxiv.org/abs/1803.08051}{{\tt
  1803.08051}}].

\bibitem{DOnofrio:2015gop}
M.~D'Onofrio and K.~Rummukainen, \emph{{Standard model cross-over on the
  lattice}}, \href{http://dx.doi.org/10.1103/PhysRevD.93.025003}{\emph{Phys.
  Rev.} {\bf D93} (2016) 025003}, [\href{http://arxiv.org/abs/1508.07161}{{\tt
  1508.07161}}].

\end{thebibliography}\endgroup

\end{document}